\documentclass[10pt,a4paper,notitlepage]{elsarticle}
\usepackage[utf8]{inputenc}
\usepackage[english]{babel}
\usepackage{amsmath,bm}
\usepackage{amsfonts}
\usepackage{amssymb}
\usepackage{graphicx}
\graphicspath{{./graphics/}}
\usepackage{subcaption}
\usepackage{color}

\newcommand*\rfrac[2]{{}^{#1}\!/_{#2}}

\begin{document}

\begin{frontmatter}

\title{Direct Numerical Simulations of pore competition in idealized 
micro-spall using the VOF method}

\author[a,b]{L.C. Malan\corref{correspondingauthor}\fnref{present_LM}} 
\fntext[present_LM]{Present address: Department of Mechanical Engineering, University of Cape Town, 
Cape Town 7701, South Africa}
\cortext[correspondingauthor]{Corresponding author}
\ead{lcmalan@gmail.com}

\author[a,b]{Y. Ling\fnref{present_YL}}
\fntext[present_YL]{Present address: Department of Mechanical Engineering, Baylor University, Waco, 
Texas 76798, United States of America}
\author[c]{R. Scardovelli}
\author[d]{A. Llor}
\author[a,b]{S. Zaleski}

\address[a]{Sorbonne Universit\'{e}s, UPMC Univ Paris 06, UMR 7190, 
Institut Jean le Rond d'Alembert, F-75005, Paris, France}
\address[b]{CNRS, UMR 7190, Institut Jean le Rond d'Alembert, 
F-75005, Paris, France}
\address[c]{DIN-Laboratorio di Montecuccolino, Universit\`a di Bologna, 
40136 Bologna, Italy}
\address[d]{CEA, DAM, DIF, 91297 Arpajon Cedex, France}
\date{}

\begin{abstract}
Under shock loading, metals have been found to melt and with reflection of the shock wave from the material free surface, cavities nucleate and grow. This process is referred to as micro spall and has been studied experimentally with analytical models describing debris sizes \cite{2008SignorThesis, 2008Signor}. 
Measurements during the cavity growth phase are not possible at present and we 
present here the Direct Numerical Simulation of an idealized problem where we assume 
an inviscid, incompressible liquid subject to a constant expansion rate with 
cavities at a vanishing vapour pressure.

In order to allow for a time-varying gas volume a free-surface interface 
condition has been implemented in an existing incompressible multiphase 
Navier-Stokes solver, PARIS Simulator, using a volume-of-fluid method. 
The gas flow remains unsolved and is instead assumed to have a fixed pressure which is applied to the liquid through a Dirichlet boundary condition on the arbitrary liquid-gas interface. 
Gas bubbles are tracked individually, allowing the gas pressure to be prescribed using a suitable equation of state.

Simulations with hundreds of bubbles have been performed in a fixed domain under 
a constant rate of expansion. A bubble competition is observed: larger bubbles 
tend to expand more rapidly at the demise of smaller ones. The time scale of 
competition is shown to depend on a modified Weber number. 
\end{abstract}


\end{frontmatter}

\section{Introduction}
Micro spall refers to failure of a material when fragmentation occurs in melted 
parts under shock loading \cite{2008Signor}. The material sample can be subject 
to projectile impact, explosive detonation or laser irradiation, events that 
in turn create a compression wave in the material. Upon reflection from 
the free surface, tensile stresses are created in the material that cause the 
nucleation of cavities that may grow up to coalescence and lead to fine droplets being 
formed. It has been studied experimentally 
\cite{2009DeResseguier,2010DeResseguier,2010aSignor} with various perspectives 
on void fraction evolution and debris sizes 
\cite{2014Fuster,2010bSignor,2008Signor,2001Stebnovskii}.
% Expand more on models

The focus of this paper will specifically be on an expanding liquid containing nucleated cavities. We are interested in studying the cavity evolution during expansion of the liquid and use a Direct Numerical Simulation (DNS) approach for this purpose. This has been an increasingly popular method of studying two-phase flows 
\cite{1999Scardovelli, 2006Tryggvason}. We also adopt an incompressible flow assumption with constant fluid density. Under these 
assumptions there is no thermodynamic relation between the pressure and fluid 
density, thus allowing mass and momentum conservation to be solved without 
requiring energy conservation to close the system. 
For multi-phase DNS in general some interface tracking or capturing method 
is then employed, which determines the fluid density and viscosity by a tracked indicator function. For more detail on DNS of multiphase flows, refer to the book by Tryggvason, Scardovelli and Zaleski \cite{2011Book}.

In our problem we cannot apply the incompressible assumption to both phases, 
as the cavitation process causes gas bubbles to change in volume. One approach 
is to indeed solve an additional energy equation and then use the temperature 
field to calculate the mass transfer at the interface between the two otherwise 
incompressible, immiscible fluids \cite{2005Tryggvason}.

Another approach is to assume an incompressible liquid, where mass and momentum 
conservation is enforced by the incompressible Navier-Stokes equations. 
The gas phase, however, is left unsolved. Instead, a free-surface condition is 
applied on the interface and the gas phase only effects the flow through its 
pressure. This is an accurate approach when the density ratio is high and 
has been used extensively. An early example is the marker code of Harlow and 
Welch \cite{1965Harlow}. More recently it has been used by Popinet to study 
bubble collapse and jet formation \cite{2002Popinet}. 
An example of an industrial application of this approach is the study of the 
ink ejection process in an inkjet printer head \cite{2016Tan}. Using a 
free-surface condition is also the approach we will use in this study, since 
the density ratios in the fluids typically present in actual micro-spall 
experiments are very large. 

\section{Mathematical Formulation}
For this study we will assume zero viscosity, motivated by the relatively small Ohnesorge number encountered in micro-spall experiments with Tin samples. Typical Tin samples have a thickness of the order of $100 \mu m$. If we were to consider the length scale of the small debris in Laser irradiation and plate impact experiments, $L \approx 10 \mu m$ \cite{2010aSignor}, we have 

\begin{equation}
Oh = \dfrac{\mu}{\sqrt{\rho \sigma L}} = 5.5 \times 10^{-3} \ \,
\end{equation} 
with $\mu = 10^{-3} \ Pa.s $, $\rho = 6.5 \times 10^{3} \ kg.m^{-3}$ and $\sigma = 0.5 \ N.m^{-1}$ the dynamic viscosity, density and surface tension of liquid Tin. 

We solve the flow in an incompressible, inviscid liquid with a sharp interface 
of arbitrary shape to a gas phase. We assume this interface moves freely and apply a Dirichlet condition for the pressure at the interface. This pressure value is obtained using the gas phase pressure as well as the pressure jump due to surface tension. The liquid flow is then governed by the incompressible Euler equations
%-----------------------------------------
\begin{equation}
\dfrac{\partial \bm{u}}{\partial t} + \bm{u} \cdot \nabla \bm{u} = 
-\dfrac{\nabla p}{\rho} \,, 
\end{equation}
%-----------------------------------------
with $\rho$ and $\bm{u}$ respectively the liquid density and velocity. $p$ is the pressure in the liquid. For incompressible fluids, mass conservation is given by
%-----------------------------------------
\begin{equation}
\nabla \cdot \bm{u} = 0 \,.
\end{equation}
%-----------------------------------------
The pressure of the gas phase is determined from an equation of state. Since
we are assuming adiabatic conditions, we consider a polytropic gas law 
\cite{1993Szymczak} to define the pressure of the gas
%-----------------------------------------
\begin{equation}\label{eq:polygas}
p_{g} = p_{0} \left(\frac{V_{0}}{V_{g}}\right)^{\gamma} \,,
\end{equation}
%-----------------------------------------
where $V_{g}$ is the total volume of the gas at pressure $p_{g}$. $p_{0}$ and $V_{0}$ are the respective reference pressure and volume of the 
gas phase  and $\gamma$ is the heat capacity ratio.

The pressure at the free surface on the liquid side, $p_{s}$, is equal to the 
gas pressure with the addition of the Laplace pressure jump due to surface 
tension
%-----------------------------------------
\begin{equation}
p_{s} = p_{g} + \sigma \kappa \, ,
\label{Laplace}
\end{equation}
where $\sigma$ is the surface tension coefficient, assumed to be constant. The interface curvature is given by $\kappa$.
%-----------------------------------------
The interface is captured using a volume-of-fluid \cite{1981Hirt} approach, 
that considers a \emph{colour} function, $c$, that obeys the following 
advection equation
%-----------------------------------------
\begin{equation} 
\dfrac{\partial c}{\partial t} + \nabla \boldsymbol{\cdot} 
\left( c \,\boldsymbol{u} \right) = 0 \,.
\label{eq:vof_advect}
\end{equation}
%-----------------------------------------
The function $c$ represents the volume fraction or volume-of-fluid (VOF) of a 
reference phase present in the spatial domain. For an arbitrary volume 
$\Omega$, $c$ is then given by
%-----------------------------------------
\begin{equation}
c (t) = \frac{1}{\Omega} \int_{\Omega} \chi \left( \bm{x},t \right) \: 
dx \, dy \, dz \,,
\end{equation}
%-----------------------------------------
where $\chi \left( \bm{x},t \right)$ is the characteristic function,
satisfying $\chi=1$ inside the reference phase and $\chi=0$ outside.

\section{Numerical Method}
The governing equations are discretized on an equi-spaced Cartesian mesh in the so-called MAC arrangement \cite{1965Harlow}. Volume-averaged scalar values ($p$ and $c$) are located in the center of computational cells, while scalar components of velocity are located on cell faces. The density is constant, since we are only considering the liquid flow.

\subsection{Time integration}
The above system of equations is solved numerically using a projection method 
\cite{1969Chorin}. The discrete form of the equations that follow are written for an explicit first order time integration to illustrate the numerical procedure. 
First, a temporary velocity field $\bm{u^{*}}$ is obtained by solving
%-----------------------------------------
\begin{equation}
\dfrac{\bm{u}^{*}-\bm{u}^{n}}{\Delta t} = 
- \bm{u}^{n} \cdot \bm{\nabla}^{h} \bm{u}^{n}
\label{eq:u_temp}
\end{equation}
%-----------------------------------------
where $\Delta t$ is the time step, the superscript $n$ refers to the $n$-{th} time step and $\bm{\nabla}^{h}$ is the discrete gradient operator. 
The velocity at the next time step, $n+1$, is then obtained by adding the 
contribution of the pressure term
%-----------------------------------------
\begin{equation}
\dfrac{\bm{u}^{n+1}-\bm{u}^{*}}{\Delta t} = -\dfrac{\bm{\nabla}^{h} p^{n+1} }{\rho} \,.
\label{eq:correct}
\end{equation}
The pressure gradient will be calculated to include surface tension at the interface. This will be detailed in section \ref{subsec:free_surface}.
%-----------------------------------------
For incompressibility of the flow we require
%-----------------------------------------
\begin{equation}
\bm{\nabla}^{h}\cdot \bm{u}^{n+1} = 0 \,,
\label{eq:non-divergence}
\end{equation}
%-----------------------------------------
and by substituting \eqref{eq:correct} in \eqref{eq:non-divergence} we have
%-----------------------------------------
\begin{equation}
\bm{\nabla}^{h} \cdot \left[\frac{\Delta t}{\rho} \bm{\nabla}^{h} p^{n+1} 
\right] = \bm{\nabla}^{h} \cdot \bm{u}^{*} \,.
\label{eq:press}
\end{equation}
%-----------------------------------------
We therefore find the divergence-free velocity field at time step $n+1$ by correcting the temporary velocity field $\bm{u^{*}}$ with the pressure found by solving \eqref{eq:press}  and then using \eqref{eq:correct}
%-----------------------------------------
\begin{equation}
\bm{u}^{n+1} =\bm{u^{*}} - \dfrac{\Delta t}{\rho}\bm{\nabla}^{h} p^{n+1} \,.
\label{eq:vel_correct}
\end{equation}
%-----------------------------------------
These equations are solved for the liquid flow. We have an arbitrary, free surface interface to gas phase and track the liquid using a VOF method, for which we solve an advection equation
%-----------------------------------------
\begin{equation}
c^{n+1} = c^{n} - \Delta t \left[ \bm{\nabla}^{h} \cdot 
\left( c \bm{u} \right)^{n} \right] \,.
\label{eq:vof_advect_time}
\end{equation}
%-----------------------------------------
This equation is solved in two steps: reconstruction of the interface as a plane in each grid cell and then its advection with the computation of the reference phase fluxes across the cell boundary. The use of planes to reconstruct the interface is accredited to de Bar \cite{1974DeBar}. In the first part of the reconstruction step, the interface normal $\bm{n}_{s}$ is computed with the “mixed Youngs-centered” (MYC) method \cite{2007Aulisa}. Then the position of a plane, representing the interface in the cell, is determined using elementary geometry \cite{2000Scardovelli}
%-----------------------------------------
\begin{equation}
\bm{n}_{s} \cdot \bm{x} = n_{sx} x + n_{sy} y + n_{sz} z = \alpha \,,
\end{equation}
%-----------------------------------------
where the scalar $\alpha$ characterizes the position of the interface. 
For the computation of the reference phase fluxes we can use the Lagrangian explicit CIAM scheme \cite{1995Li} or the strictly conservative Eulerian scheme of Weymouth and Yue \cite{2010Weymouth}. 

\label{sec:paris_std}

A well-known method \cite{2004Esmaeeli, 2003Sussman} to increase the convergence order of time integration is to calculate two explicit time steps and halve the result

\begin{align}
u^* =& u^{n} + \tau L(u^{n}) \\
u^{**} =& u^* + \tau L(u^{*}) \\
u^{n+1,*} =& \frac{1}{2} (u^{**} + u^n )
\end{align}

%-----------------------------------------
\subsection{Treatment at the free surface}
\label{subsec:free_surface}
In general the method to deal with VOF advection and surface tension is similar to \textsc{Gerris} \cite{2009Popinet}. This section will describe the treatment of the interface as a free surface.

At the interface to the gas phase, we need to apply a Dirichlet condition for the pressure to include the effects of the gas pressure and surface tension on the liquid flow. The method used in this work is inspired by the idea of Fedkiw and Kang \cite{1999Fedkiw,2000Kang}, often referred to as the ghost fluid method. First, we need to find the gas pressure from 
\eqref{eq:polygas}. In this equation, $p_{0}$ and $V_{0}$ are known gas 
quantities so the gas volume needs to be determined.
 
This is done by identifying continuous volumes of gas inside the domain using the colour function, $c$ and a numerical algorithm based on the work by Herrmann \cite{2010Herrmann}. A viral tagging procedure is used to mark computational cells containing the desired phase, after which connected cells are agglomerated into a single volume. The procedure is compatible with domain decomposition in parallel processing.

With the value of $p_{g}$ calculated for each gas bubble, we need to 
discretize \eqref{eq:press} for liquid cells near the interface. 
Cells that contains mostly gas are excluded from the solution, so we only solve for nodes where $c<0.5$, with $c=0$ in the liquid.  

\begin{figure} 
  \centering
  \def\svgwidth{0.6\columnwidth}
  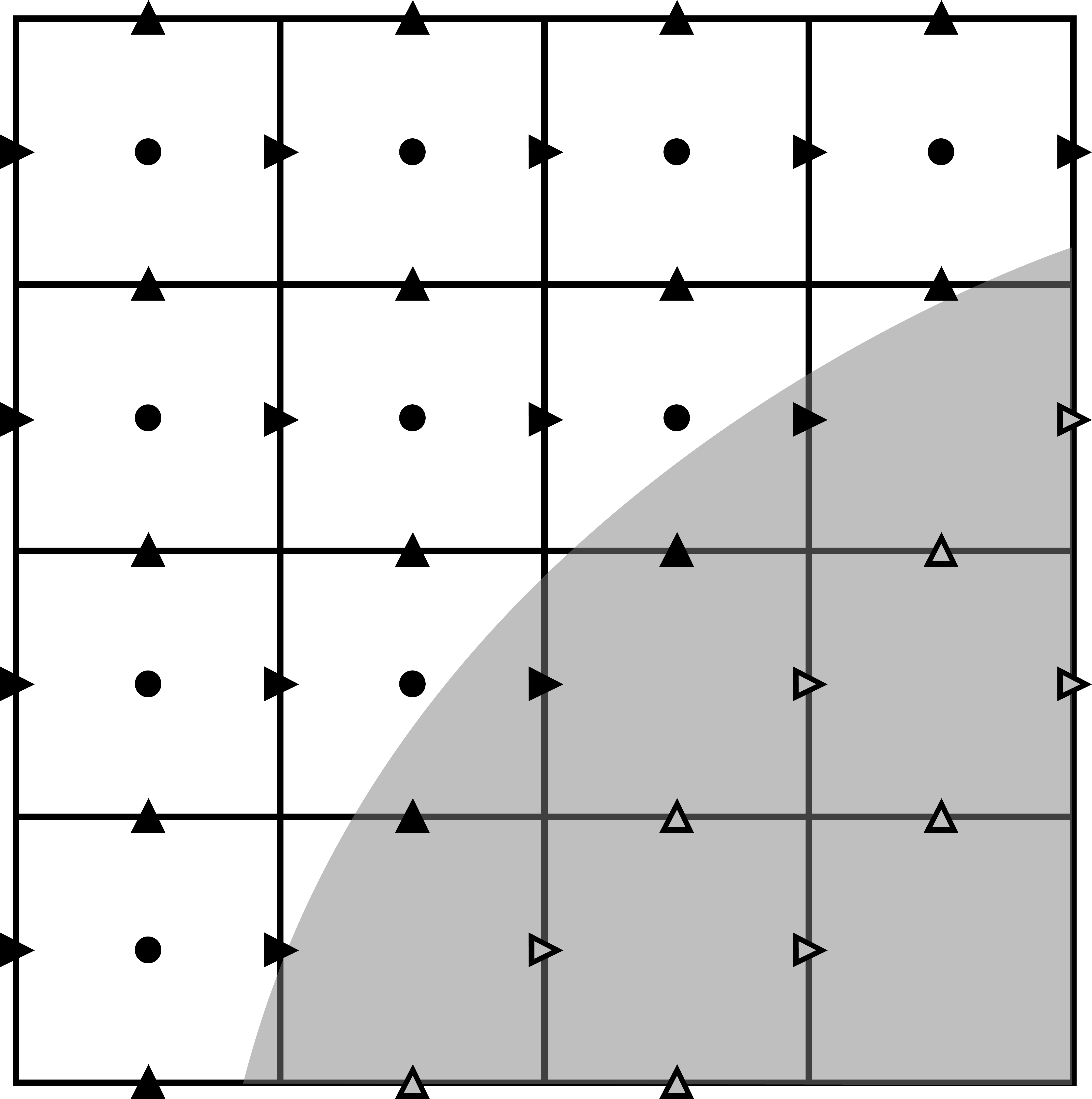
  \caption{A 2D section of the numerical grid, showing part of a gas bubble in grey. Circles represent computational cell nodes, where pressure is calculated. Triangles indicate scalar velocity components on the computational cell faces. Filled triangles indicate values which are found by solving the governing equations, while unfilled triangles represent boundary values found by extrapolation.}
  \label{fig:p_nodes}
\end{figure}

\begin{figure}
  \centering
  \begin{subfigure}[b]{1.00\textwidth}
    \def\svgwidth{0.825\textwidth}
    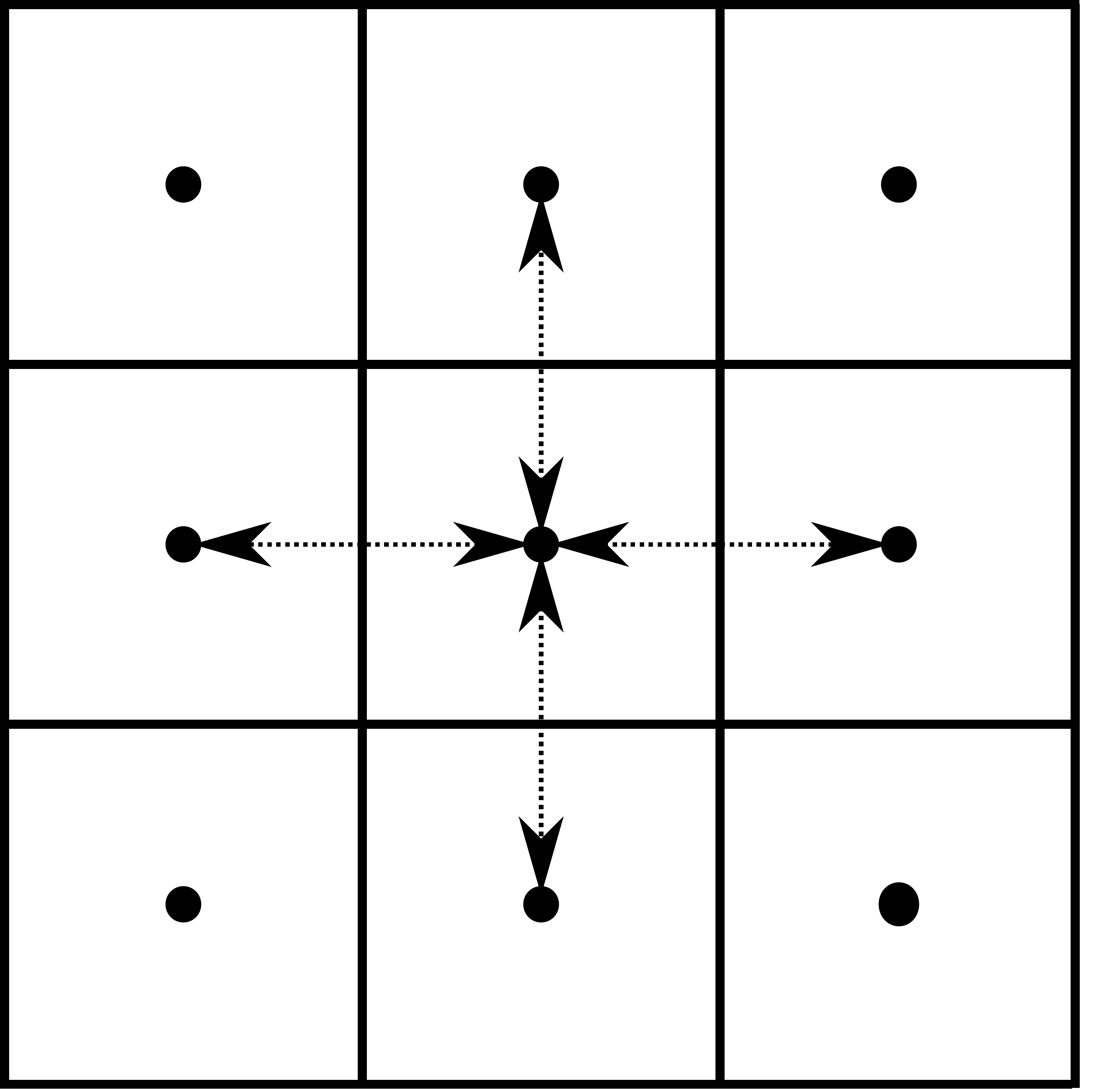
    \caption{Standard discretisation of the pressure equation 
             in the liquid bulk}
    \label{fig:p_bulk_branches}
  \end{subfigure}
  \begin{subfigure}[b]{1.00\textwidth}
    \def\svgwidth{0.80\textwidth}
    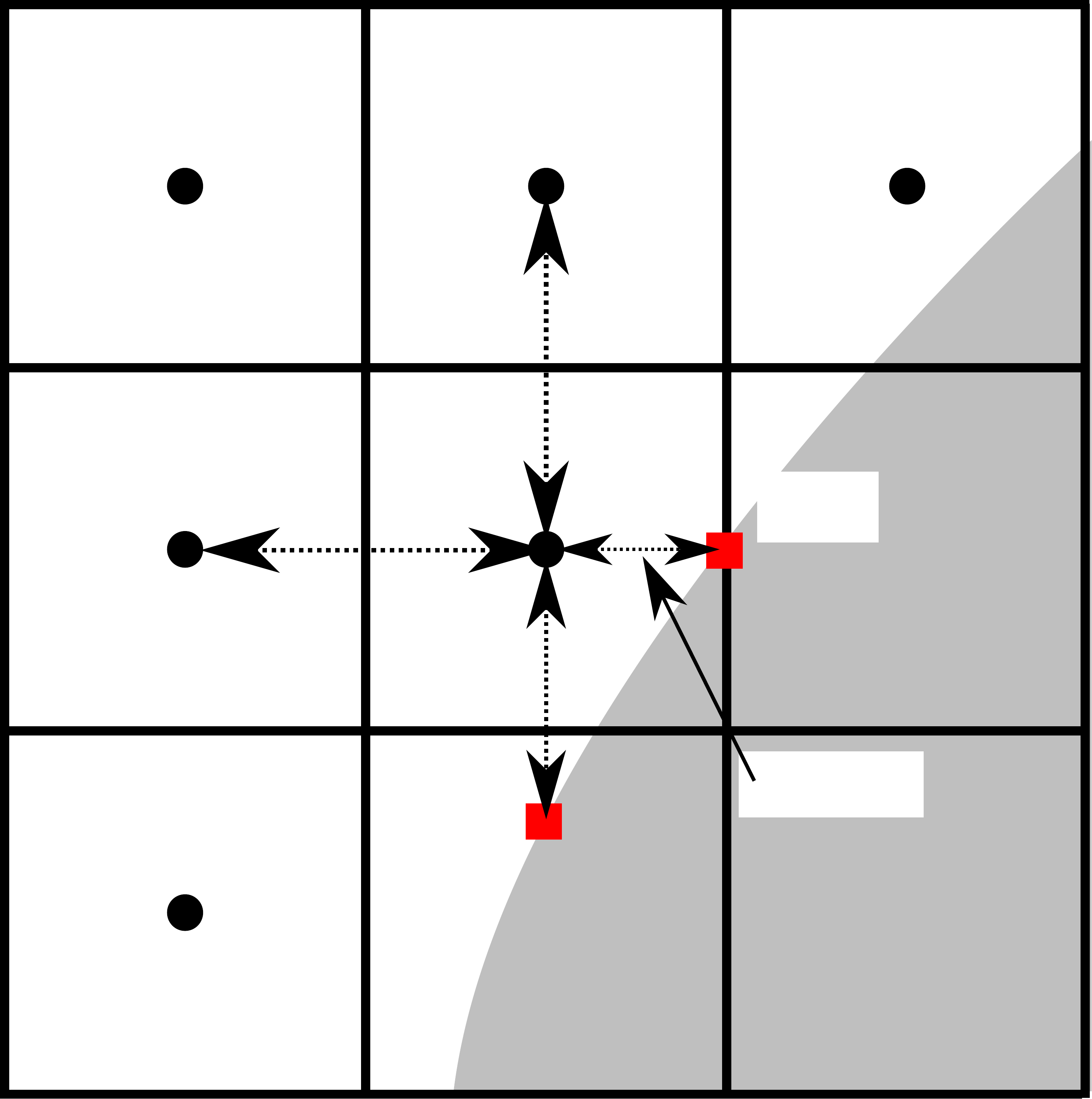
    \caption{Discretisation of the pressure equation near the interface}
  	\label{fig:p_mod_branches}
  \end{subfigure}
\end{figure}

Fig. \ref{fig:p_nodes} shows a representation of a 2D grid with a 
section of an interface. The grey area represent a vapour-filled cavity. Cells that contain a filled circle are included in the pressure solution, while cells without a marker are excluded.
%-----------------------------------------
%-----------------------------------------
We first provide the finite volume discretisation of the left hand side 
of \eqref{eq:press} for a bulk liquid cell in 2D, shown in Fig. \ref{fig:p_bulk_branches}.
%-----------------------------------------
\begin{align}
\frac{\Delta t}{V_{i,j}} & \int_{V_{i,j}} \bm{\nabla} \cdot 
\left[\frac{\bm{\nabla}^{h}\, p^{n+1}}{\rho} \right] dV 
\nonumber \\
\approx & \dfrac{\Delta t}{\rho} \left( \dfrac{\nabla^{h}_y \,p_{i,j+\rfrac{1}{2}} 
- \nabla^{h}_y \,p_{i,j-\rfrac{1}{2}}}{\Delta y} + \dfrac{\nabla^{h}_x \, 
p_{i+\rfrac{1}{2},j} - \nabla^{h}_x \,p_{i-\rfrac{1}{2},j}}{\Delta x} \right) 
\nonumber \\
=& \dfrac{\Delta t}{\rho} \dfrac{p_{i,j+1} + p_{i,j-1} + p_{i+1,j} + p_{i-1,j} 
- 4 p_{i,j}}{h^{2}} \,, 
\label{pr_lhs}
\end{align}
%-----------------------------------------
where the $i$ and $j$ subscripts are integer indices for the discrete 
computational cell with volume $V_{i,j}$. We consider a constant density $\rho$ for the liquid phase. Furthermore, with cubic cells we have $\Delta x = \Delta y =\Delta z = h$, where $h$ is the constant grid spacing.

%-----------------------------------------
%-----------------------------------------
The stencil for the pressure gradient components has to be changed near the interface when a neighbouring pressure in expression \eqref{pr_lhs} falls inside the gas phase. This point must be disregarded and its pressure substituted by a surface pressure. We apply the same approach as Chan \cite{1970Chan}. As an example, we provide the approximation for the pressure gradient components for the cell with indices $i$ and $j$ in Fig. \ref{fig:p_mod_branches}
%-----------------------------------------
\begin{equation} \label{eq:mod_branches}
\bm{\nabla}^{h}_x \,p_{i+\rfrac{1}{2},j} = \dfrac{p_{s,i+1,j} -
p_{i,j}}{\delta_{i+\rfrac{1}{2},j}} \,\,;\quad
\bm{\nabla}^{h}_y \,p_{i,j-\rfrac{1}{2}} = \dfrac{p_{i,j} -
p_{s,i,j-1}}{\delta_{i,j-\rfrac{1}{2}}} \, ,
\end{equation}
%-----------------------------------------
where $\delta$ is the distance between the pressure node under consideration and the intersection with the interface. The pressure $p_{s}$ on the liquid side of the interface is found by adding to $p_{g}$ the Laplace pressure jump. The pressure $p_{g}$ inside each gas bubble is known from \eqref{eq:polygas}. The interface pressure in the x-direction will then be
\begin{equation}
p_{s,i+1,j} = p_{g,i+1,j} + \sigma \frac{\kappa_{i,j} + \kappa_{i+1,j}}{2} \,.
\label{Laplace_num}
\end{equation}
%-----------------------------------------
From \eqref{Laplace_num} and \eqref{eq:mod_branches} it is clear that accurate interface curvature as well as knowledge about its location are important parameters to ensure the accuracy of the pressure solution.

The interface curvature is computed with the height function method in a way similar to that implemented in the \textsc{Gerris} code \cite{2009Popinet}. The height function is an approximate distance to the interface from a reference cell node and is calculated by summing the cell VOF values in a column aligned with one of the principal coordinate directions, called a height stack. The principal curvature can then be obtained by using finite difference approximations for the first and second derivatives of the height function. This method has been shown to produce second order accuracy for the curvature \cite{2009Popinet}. 

It is not always possible to find all the required heights to calculate a curvature. In this case a parabolic fit is made through the plane centroids of interface cells, which is then used to estimate the curvature.

\begin{figure} 
  \centering
  \def\svgwidth{0.6\columnwidth}
  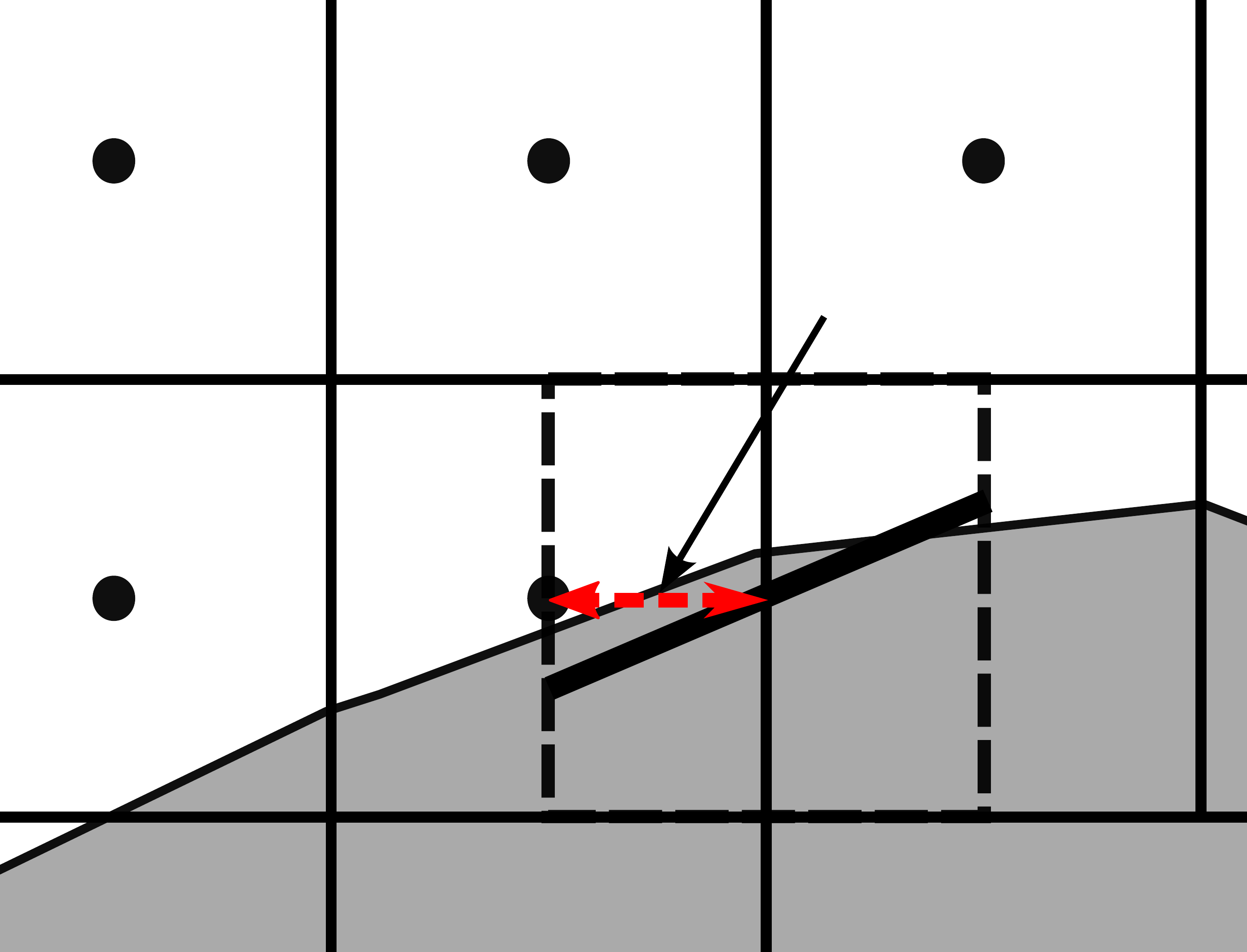
  \caption{Cell $i,j$ will typically not have a height available due to the interface configuration. A plane reconstruction (thick black line) is made in the staggered volume indicated with dashed lines and this reconstruction is used to obtain $\delta_{i+1/2,j}$.}
  \label{fig:staggered_rec}
\end{figure}

%-----------------------------------------
Since the height function is the approximate interface distance from some reference cell in a given direction, it is used for $\delta$. When the interface configuration is such that a height cannot be obtained in the required direction, the distance is approximated by using a plane reconstruction of the interface in the staggered volume. This is shown in Fig. \ref{fig:staggered_rec}.
First, the staggered VOF fractions are obtained by considering the plane reconstruction in centered cells. A similar procedure is then used in the staggered cells than in the centered cells to reconstruct the interface as a plane. With the plane constant known, the interface distance is then calculated.

The finite difference discretization of the left hand side of \eqref{eq:press} for cell $i,j$ in Fig. \ref{fig:p_mod_branches} will then be
%-----------------------------------------
\begin{align}\label{pr_lhs_mod}
\dfrac{\Delta t}{\rho} & \bm{\nabla} \cdot \left[\bm{\nabla}^{h}\, p^{n+1} \right] \nonumber \\
&\approx \dfrac{\Delta t}{\rho} \left( \dfrac{\nabla^{h}_y \,p_{i,j+\rfrac{1}{2}} - \nabla^{h}_y \,p_{i,j-\rfrac{1}{2}}}
{ \rfrac{1}{2} \left( \Delta y_{j+\rfrac{1}{2}} +\Delta y_{j-\rfrac{1}{2}} \right) } 
+ \dfrac{\nabla^{h}_x \, p_{i+\rfrac{1}{2},j} - \nabla^{h}_x \,p_{i-\rfrac{1}{2},j}}{ \rfrac{1}{2} \left( \Delta x_{i+\rfrac{1}{2}} + \Delta x_{i-\rfrac{1}{2}} \right)} \right) \nonumber \\
&= \dfrac{\Delta t}{\rho} \left( \quad \dfrac{2}{h+\delta_{i,j-\rfrac{1}{2}}} \left( \dfrac{p_{i,j+1} - p_{i,j}}{h} - \dfrac{p_{i,j} -p_{s,i,j-1}}{\delta_{i,j-\rfrac{1}{2}}} \right) \right. \nonumber \\
& \qquad \qquad + \left. \dfrac{2}{\delta_{i+\rfrac{1}{2},j} + h} \left( \dfrac{p_{s,i+1,j} - p_{i,j}}{\delta_{i+\rfrac{1}{2},j}} - \dfrac{p_{i,j} - p_{i-1,j}}{h} \right) \quad \right) \,. 
\end{align}

The implementation in 3D is included in \textit{PARIS}.
%-----------------------------------------
\subsection{Extrapolation of the velocity field} 
\label{sec:velocity_extrapolation}

The previous section dealt with the treatment of the pressure at the interface. The solution of the pressure Poisson equation, \eqref{eq:press} is used in \eqref{eq:vel_correct} to correct the predicted velocities obtained in \eqref{eq:u_temp}. This section will deal with the velocity field required for the momentum contribution on the right hand side of \eqref{eq:u_temp}. The term $\bm{u}\cdot\bm{\nabla}\bm{u}$ is discretized using a choice of schemes, including QUICK \cite{1979Leonard}, ENO \cite{1986Harten} and WENO \cite{2009Shu}.

For all these schemes, the discretization of $\bm{u}\cdot\bm{\nabla}\bm{u}$ may require a velocity stencil including neighbours up to two grid spacings away, depending on the upwind direction. The discrete pressures included in the solution have been explained previously, but since we are on a staggered grid, we need to do the same for velocity components. The velocities included are all those which are on a face that has a resolved pressure directly neighbouring it in any direction. Otherwise stated, all velocities which have a pressure gradient associated with it will be resolved. These are all velocity components that are marked with filled markers in Fig. \ref{fig:p_nodes}.

As mentioned earlier, the resolved velocity components right next to the interface will require neighbours in the gas phase to discretize the momentum advection term. These values in the gas phase can be seen as boundary values to the resolved velocities. In order to find neighbours in the gas phase, we extrapolate the resolved velocities similarly to Popinet \cite{2002Popinet}.

After calculating $\bm{u}^{n+1}$ in \eqref{eq:correct}, we have a field of resolved velocities. To find the boundary velocities for the next time step, the closest two velocity neighbours inside the gas are extrapolated from the field of resolved liquid velocities using a linear least square fit. Let's assume the
velocity field can be described as a linear combination
\begin{equation}
\boldsymbol{u}\left(\boldsymbol{x}\right) = \boldsymbol{A}
\cdot (\boldsymbol{x} - \boldsymbol{x_{0}} ) + \boldsymbol{u_{0}}
\end{equation}
where the components of the tensor $\boldsymbol{A} = \nabla \boldsymbol{u}$
and of the vector $\boldsymbol{u_{0}}$ are the unknowns.\\
If we now take a $5 \times 5$ stencil around the unknown gas velocity at location
$\boldsymbol{x_{0}}$, we can find the extrapolated velocity $\bm{u_{0}}$
by minimizing the functional
\begin{equation}
\mathcal{L} = \sum_{k=1}^{N} \big | \boldsymbol{A} \cdot
(\boldsymbol{x_{k}} - \boldsymbol{x_{0}} ) + \boldsymbol{u_{0}} -
\boldsymbol{u_{k}} \big |^{2}
\end{equation}

This is done first for all locations closest to the resolved velocities
$\boldsymbol{u_{k}}$ (``first neighbours''), whereafter it is repeated
for the ``second neighbours''. Note that only resolved velocity components
are included in the cost function, therefore the number $N$ can vary depending
on the shape of the interface. Furthermore, because of the staggered grid,
only one velocity component of $\boldsymbol{u_{0}}$ is computed at
location $\boldsymbol{x_{0}}$. \\

%-----------------------------------------
\subsection{Ensuring volume conservation}\label{sec:2nd_projection}

The extrapolation of liquid velocities into the gas phase was explained in the previous section. An additional step is required to ensure that the extrapolated 
velocities are divergence free. This is required to ensure that the advection of the colour function \eqref{eq:vof_advect} is conservative.

\begin{figure} 
  \centering
  \def\svgwidth{0.6\columnwidth}
  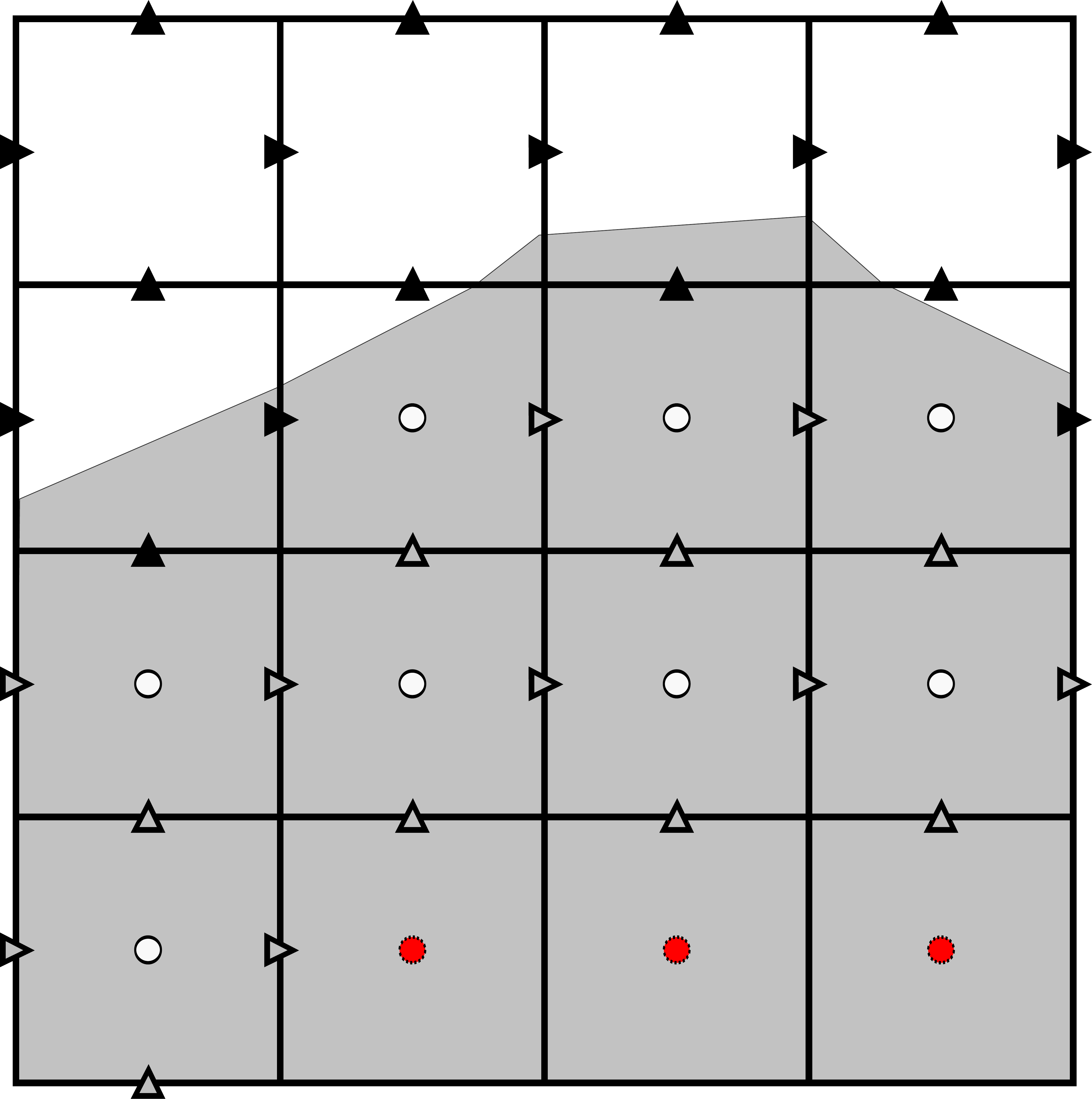
  \caption{2D example of the problem to correct the extrapolated velocities (unfilled triangles). A Poisson problem is solved in the cells marked with an unfilled circle.}
  \label{fig:2nd_proj}
\end{figure}

A similar approach to Sussman \cite{2003Sussman} is used. Only the first two layers of cells inside the gas phase are 
considered and all other cells are disregarded. A 2D example is presented in Fig. \ref{fig:2nd_proj}.
Similar to the projection step explained earlier, a ``phantom'' pressure is obtained in these cells by solving a Poisson equation 
%-----------------------------------------
\begin{equation}
\nabla^{h} \boldsymbol{\cdot} \bigg ( \nabla^{h} \hat{P} \bigg ) = 
\nabla^{h} \boldsymbol{\cdot} \boldsymbol{\tilde{u}} \,,
\end{equation}
%-----------------------------------------
where $\hat{P}$ is the ``phantom'' pressure and $\boldsymbol{\tilde{u}}$ 
is the velocity on the faces of the first two gas neighbours. $\hat{P}$ is only calculated in the cells represented by unfilled nodes in Fig. \ref{fig:2nd_proj}. On the liquid side of these cells, the solved velocities (filled triangles) are used as a velocity boundary condition with the pressure gradient on this face set to zero. On the gas side outside the cells we consider (red filled circles), a fixed pressure is prescribed. Only the extrapolated velocities (unfilled triangles) are then corrected by the solved pressure gradient, $\nabla \hat{P}$ 
%-----------------------------------------
\begin{equation}
\boldsymbol{\tilde{u}}^{n+1} = \boldsymbol{\tilde{u}} - \nabla^{h} \hat{P}
\end{equation}
%-----------------------------------------
to ensure non-divergence of velocity in the first two layers of cells just inside the gas.

%-----------------------------------------
\subsection{Rayleigh-Plesset equation and single bubble test} 
\label{sec:RP}

%-----------------------------------------
\begin{figure}[ht]
\centering
\includegraphics[width=0.7\textwidth]{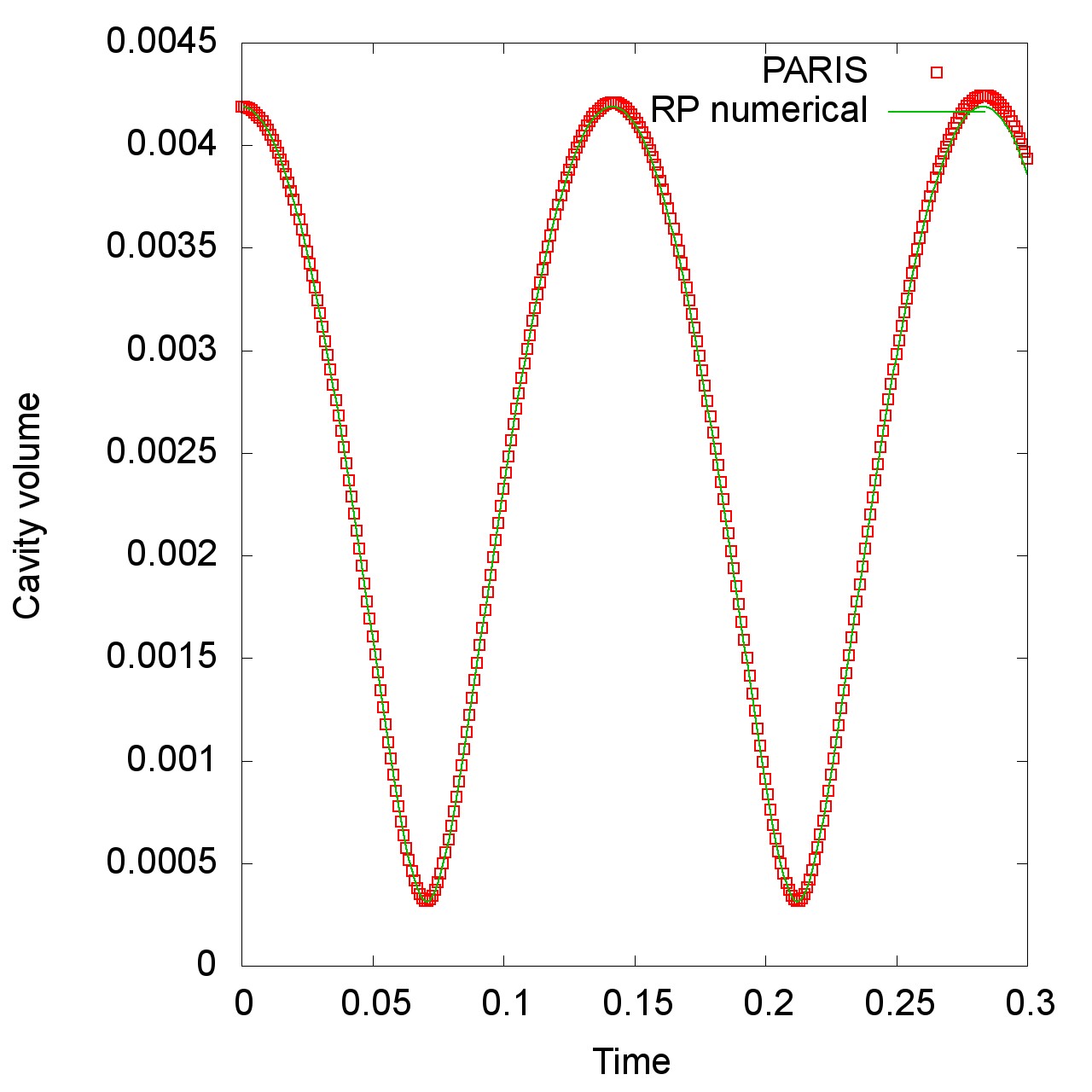}
\caption{Comparison of results of a single oscillating gas bubble 
simulated by PARIS and the Rayleigh-Plesset equation.}
\label{fig:rp}
\end{figure}
%-----------------------------------------
This section presents the validation of the numerical implementation of our 
model. A widely-used approach is to compare a numerical simulation of a single 
gas bubble with a fixed liquid pressure at infinity to the solution of the 
Rayleigh-Plesset equation \cite{1977Plesset}. This equation describes the 
evolution of a bubble of radius $R$ in an incompressible liquid, assuming 
spherical symmetry with a fixed pressure at infinity. Neglecting viscous 
effects, it is given by
%-----------------------------------------
\begin{equation} \label{eq:RP}
\begin{split}
\ddot{R}R + \dfrac{3}{2}\dot{R}^{2} &= \dfrac{p_{s} - p_{\infty}}{\rho_{l}} \\
 & = \dfrac{p_{g} - \frac{2\sigma}{R} - p_{\infty}}{\rho_{l}}
\end{split}
\end{equation}
%-----------------------------------------
where $R$ is the bubble radius, $p_{s}$ the pressure on the liquid side of the 
interface, $p_{\infty}$ the pressure at infinity, $\sigma$ the surface 
tension coefficient and $\rho_{l}$ the liquid density. 
A bubble of initial radius $R(t=0)=$0.10 is placed in a liquid with density 
$\rho_{l}=1.0$ and a surface tension coefficient $\sigma=0.10$. 
The bubble's reference pressure is $p_0=1.0$ at a reference radius  
$R_0=0.09$ and the pressure at infinity is $p_{\infty}=0.5$. 
The bubble pressure, $p_{g}$, is obtained from a polytropic gas law 
%-----------------------------------------
\begin{equation}
p_{g} = p_{0} \left(\frac{R_{0}}{R}\right)^{3 \gamma}
\end{equation}
%-----------------------------------------
where $\gamma=1.4$ is the isentropic gas coefficient. The domain used for the 
simulation is a cube of size 1.0, with the bubble placed exactly at its center
and a grid resolution of 128 cells per coordinate direction.

In the Rayleigh-Plesset equation $p_{\infty}$ is the pressure at infinity. 
However, in our numerical simulation we have to apply a pressure condition 
on the boundary of the physical domain, which is at some finite distance from the 
bubble center. The pressure to be applied at the boundary is found by solving 
numerically the Rayleigh-Plesset equation  with a 5th order 
Runge-Kutta integration method and at discrete time steps that coincide with
those of the PARIS simulation. We assume a bubble that is initially stationary, $\dot{R}(t=0) = 0$.

The solution of \eqref{eq:RP} then gives us $R(t)$ and $\dot{R}(t)$, 
that can then be used to find the pressure at some finite radius $r$ 
%-----------------------------------------
\begin{align}
\begin{split}
  &p(r,t) = p_{s} \ - \\
  &\rho_{l} \left(\frac{\dot{R^{2}}R^{4}}{2r^{4}} - 
  \frac{\ddot{R}{R}^{2}+2R\dot{R^{2}}}{r} + \ddot{R}R + \frac{3}{2}\dot{R^{2}} 
  \right)
\end{split}
\end{align}
%-----------------------------------------
hence the pressure distribution is not constant on the cube boundary.
Fig. \ref{fig:rp} shows a comparison between the results in PARIS and a 
numerical solution of the Rayleigh-Plesset equation. A good comparison 
was achieved. 
%\review{Convergence study can be performed on the boundary to interface distance, but will take significant time. Is it possible that we could get by without it? Maybe if reviewers require this?}
%-----------------------------------------

\section{Results}
%-----------------------------------------
\subsection{Non-dimensional numbers for the multiple bubble tests}

We now consider a computational domain with multiple bubbles inside 
and define a number of non-dimensional parameters to classify the flow.
The only relevant physical parameters of the fluid are its density and 
surface tension. With these parameters we can define a capillary
time scale $\tau_{R}$ using the bubble radius $R$ as the reference length
%-----------------------------------------
\begin{equation}\label{eq:cap_R}
\tau_{R} = \left( \dfrac{\rho R^{3}}{\sigma} \right) ^ { \frac{1}{2} }
\end{equation}
%-----------------------------------------
Alternatively, we could also use the inter-bubble distance as the 
characteristic length scale
%-----------------------------------------
\begin{equation}
\tau_{\ell_{D}} = \left( \dfrac{\rho \ell_{D}^{3}}{\sigma} 
\right)^{\frac{1}{2}}
\end{equation}
%-----------------------------------------
where $\ell_{D}$ is the mean bubble separation distance for $N$ bubbles 
in a cubic domain of side $L$ and is given by the expression 
%-----------------------------------------
\begin{equation}
\ell_{D} = \left( \dfrac{L^{3}}{N} \right) ^ {\frac{1}{3}}
\end{equation}
%-----------------------------------------
Furthermore, the expansion rate $\omega$ of the flow for a constant normal 
outflow velocity $U_{n}$ can be defined as
%-----------------------------------------
\begin{equation}
\omega = \frac{6 U_{n}}{L}
\end{equation}
%-----------------------------------------
Finally, we can now construct a Weber number based on the mean bubble 
separation distance
%-----------------------------------------
\begin{equation}
We_{\ell_{D}} = \frac{\rho \ell_{D}^{3} \omega^{2}}{\sigma}
\end{equation}
%-----------------------------------------
\subsection{Simulation setup}

\begin{figure}[ht]
	\centering
 	\def\svgwidth{0.6\columnwidth}
 	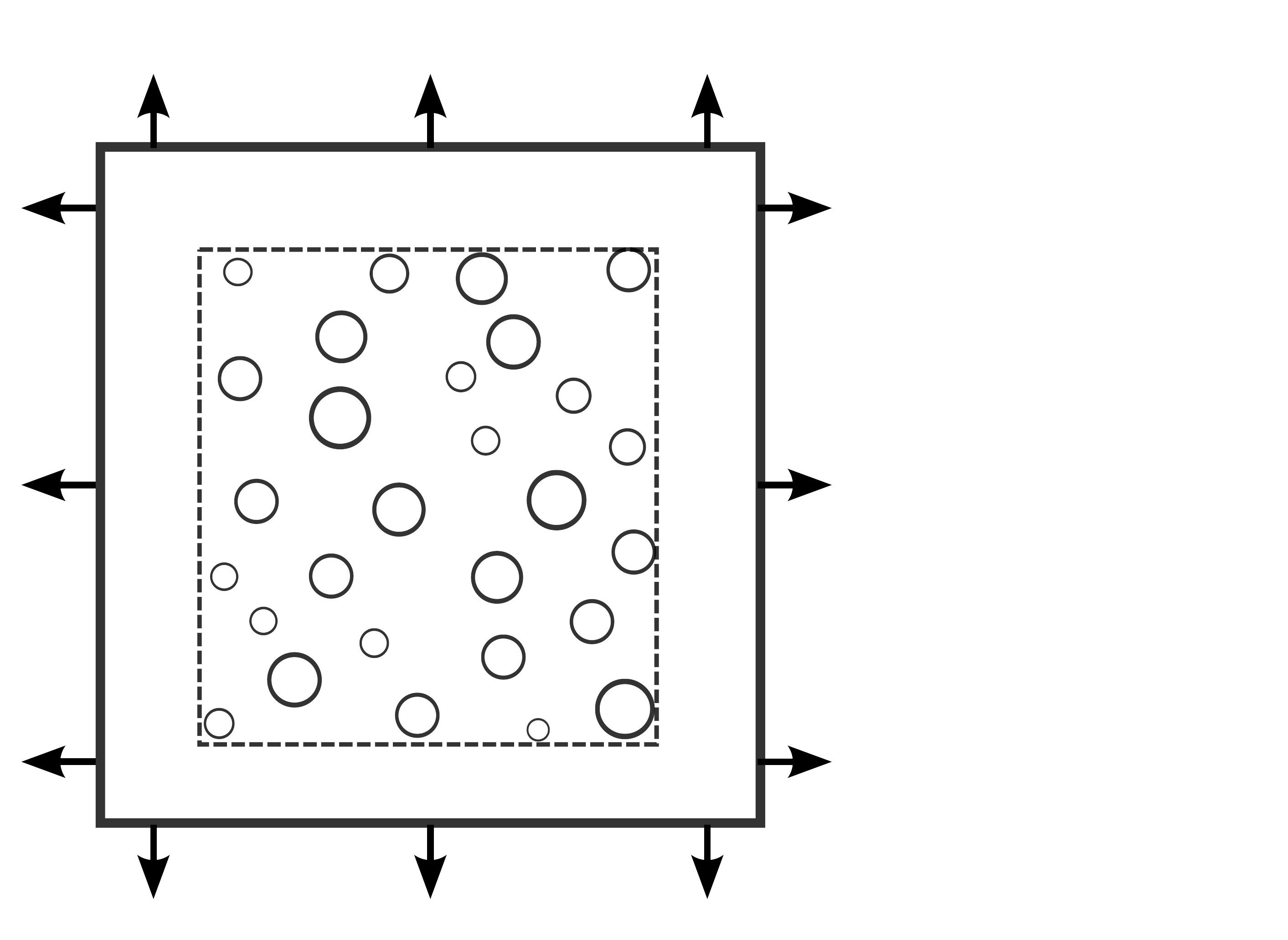
	\caption{2D slice through domain showing typical simulation setup. A uniform velocity outflow rate is specified on the domain faces. Bubbles are initialised in an internal bubble zone, surrounded by a layer of pure liquid.}
	\label{fig:sim_setup}
\end{figure}

For the following tests a constant expansion rate $\omega$ is used, hence
a constant outflow velocity $U_n$ is imposed on all the faces of the cubic domain. 
As mentioned earlier, the simulations are started with bubbles already 
at finite size. Bubbles are seeded in a face centered cubic (FCC) lattice. 
The bubble positions can correspond to the exact lattice nodes, or with some 
random displacement around this position. Fig. \ref{fig:sim_setup} shows a 
2D slice of a typical simulation setup.
%-----------------------------------------
%-----------------------------------------
Bubbles are placed in a central zone, referred to as the bubble zone. 
An all-liquid buffer zone borders the 
bubbles. The size of this zone is chosen conservatively such that only 
liquid exits the domain up to a void fraction of approximately $30\%$.  
The inter-bubble distance $\ell_{D}$ is determined by the specified number 
of FCC cells and the size of the bubble zone.

Table \ref{table:params} gives the simulation parameters used for a series 
of tests. 
%-----------------------------------------
\begin{table}
\centering
\begin{tabular}[width=0.40\textwidth]{ll}
  \hline
  \textbf{Parameter} & \textbf{Value}
  \\ \hline
  Buffer to domain length ratio & $0.12$
  \\ \hline
  Expansion rate $\omega$ & $0.033,\; 0.165,\; 1.05$
  \\ \hline
  Initial bubbles $N_0$ & $ 365 $
  \\ \hline
  Grid points & $512^{3}$
  \\ [2pt] \hline
  $We_{\ell_{D}}$ & $5 \cdot 10^{-4}, \; 0.013, \; 0.54$\\[2pt]
  \hline
  $\Delta R_{0} / R_{0}$ & $0.5$ \\[2pt]
  \hline
  $\Delta \ell_{D} / \ell_{D}$ & $0, \; 0.2$ \\[2pt]
  \hline
  $\ell_{D} / R_{0}$ & $10, \; 20$\\	[2pt]
  \hline
\end{tabular}
\caption{Simulation parameters for multiple bubble tests.}
\label{table:params}
\end{table}
%-----------------------------------------
Bubbles are initialized with an initial radius $R_{0}$ and the parameter 
$\Delta R_{0}/R_{0}$ describes the variance in the initial bubble 
diameter $R_{0}$. Bubbles are initialized with a random radius such that 
$R_{min} < R_{0} < R_{min} + \Delta R_{0}$.

Once the center position and radius of each bubble have been generated,
the colour function field can be easily and accurately initialized with the 
\textsc{Vofi} library \cite{2016Bna}.

%-----------------------------------------
\subsection{Multiple bubbles in a liquid under constant expansion}

In this section the results of three test cases with $N_0=365$ initial bubbles 
in an expanding domain will be given. By varying the normal outflow velocity 
$U_n$, different expansion rates, $\omega$ are obtained. Three velocities are considered,
$5.5 \times 10^{-3}$, $2.75 \times 10^{-2}$ and $1.75 \times 10^{-1}$,
corresponding to the three Weber numbers, $5 \times 10^{-4}$, 
$1.3 \times 10^{-2}$ and $0.54$, of Table \ref{table:params}.

\begin{figure}[h]
\centering
\includegraphics[width=0.7\textwidth]{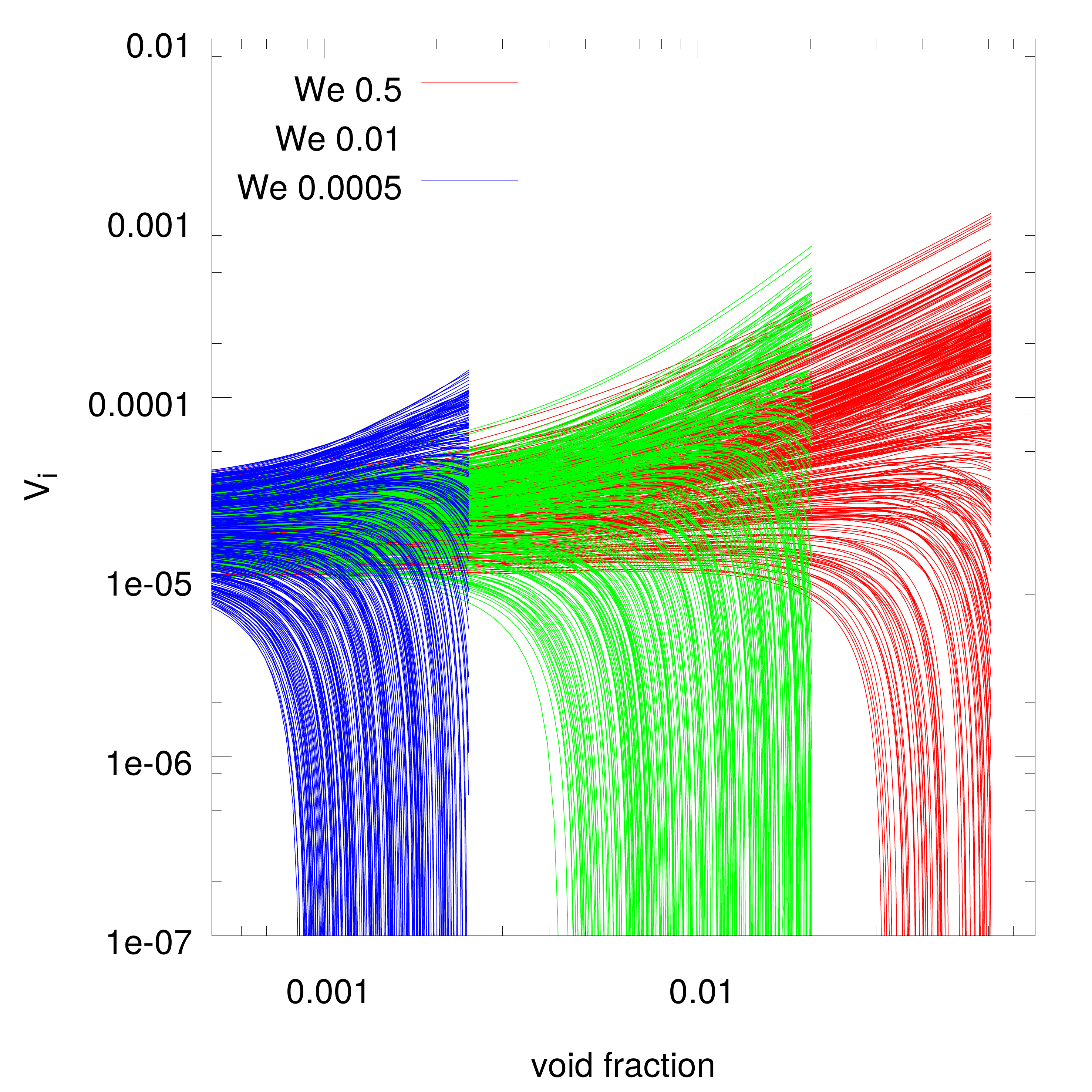}
\caption{Comparison of individual bubble volumes for varying We. Bubble 
collapse is delayed with increasing We as the domain expansion counters
capillary effects. Individual bubble volumes are plotted as function of 
total void fraction.}
\label{fig:bub_vols_we}
\end{figure}

The effect of We on the simulation results can be appreciated in Fig. 
\ref{fig:bub_vols_we}, where individual bubble volumes are plotted against 
the total void fraction. Since we are using constant outflow rates, the void fraction is directly proportional to time.
 
\begin{figure*}%[ht]
\begin{tabular}[width=1.05\textwidth]{cc}
\includegraphics[width=.5\textwidth]{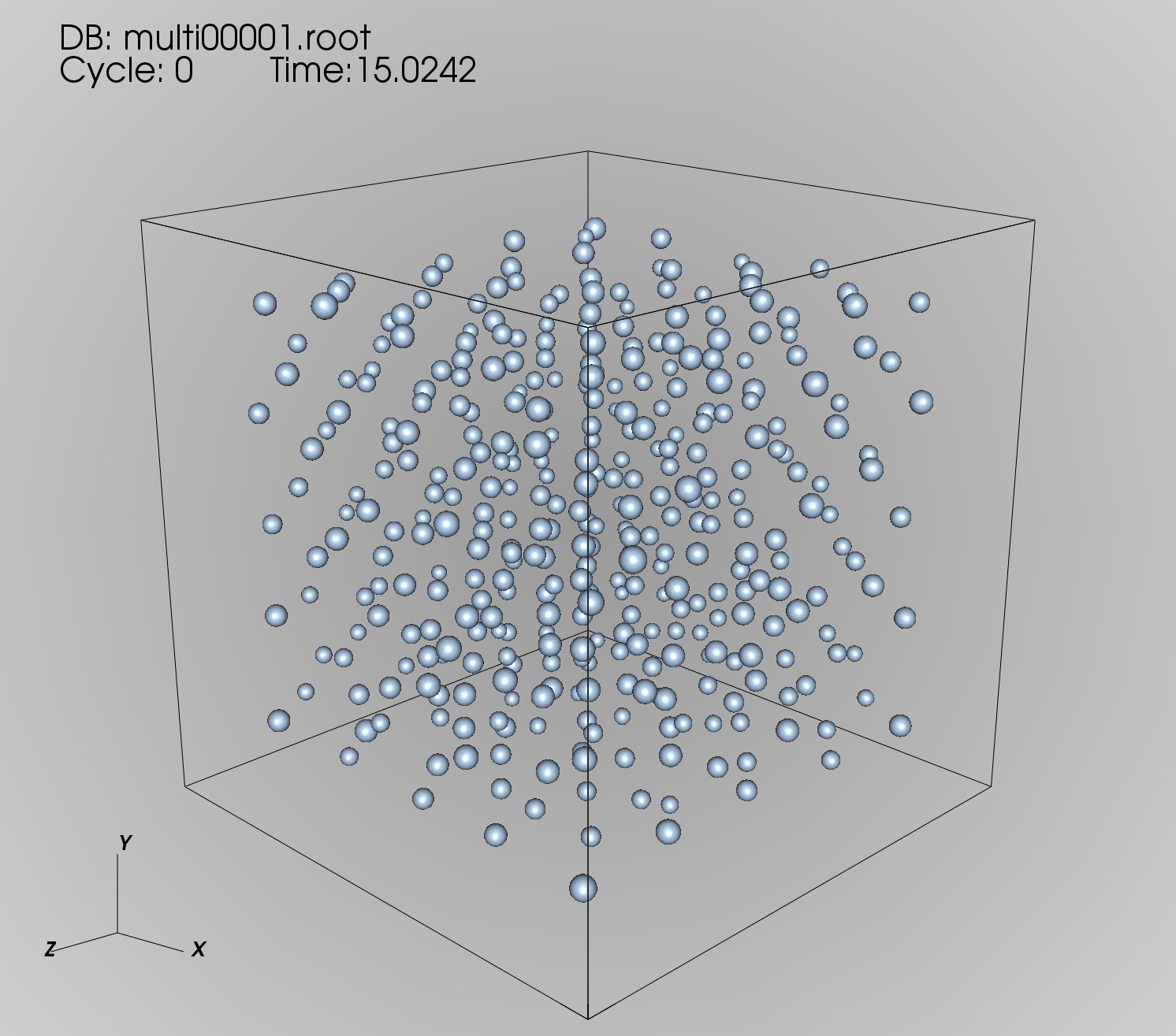} & 
\includegraphics[width=.5\textwidth]{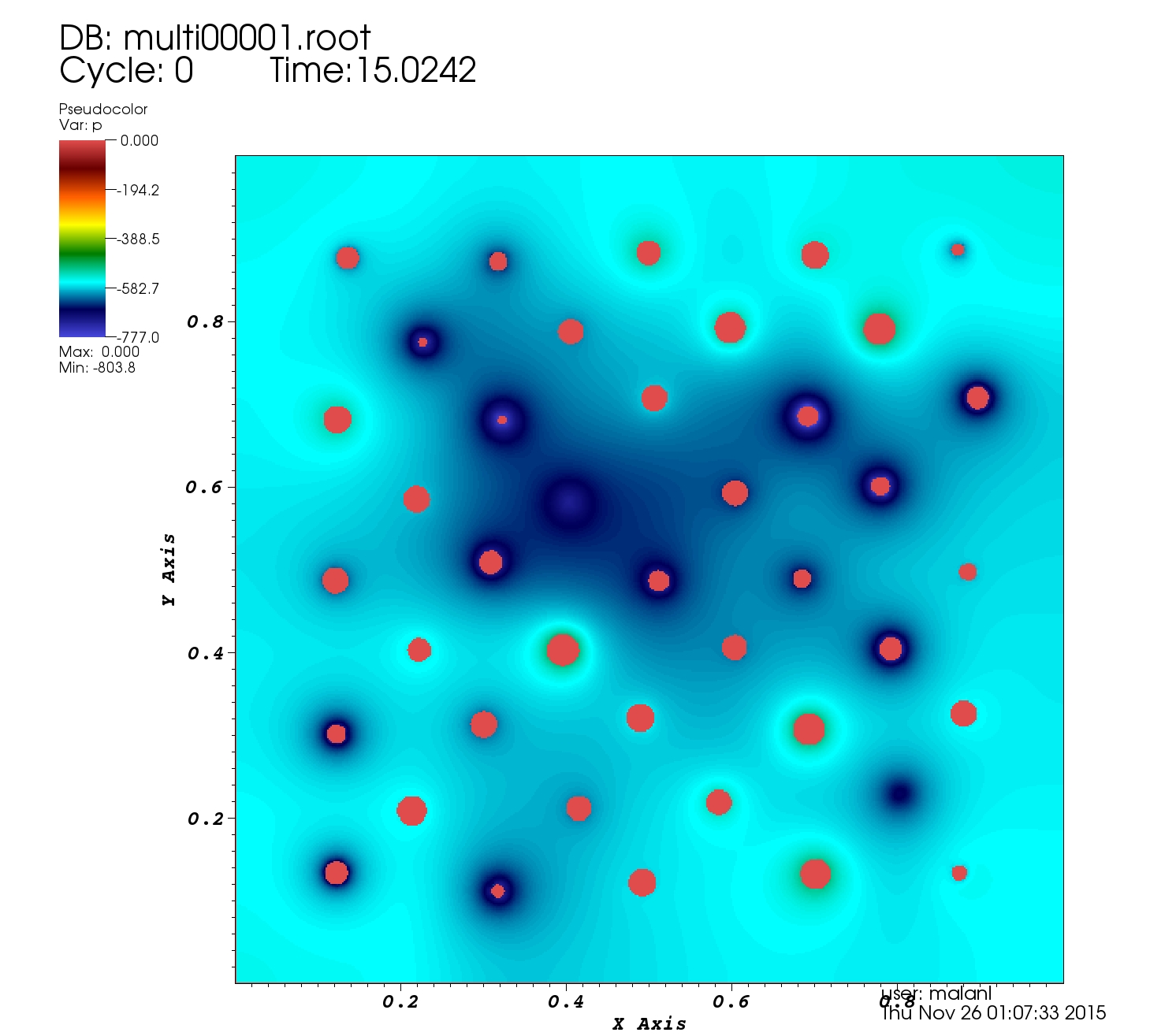} \\ 
\includegraphics[width=.5\textwidth]{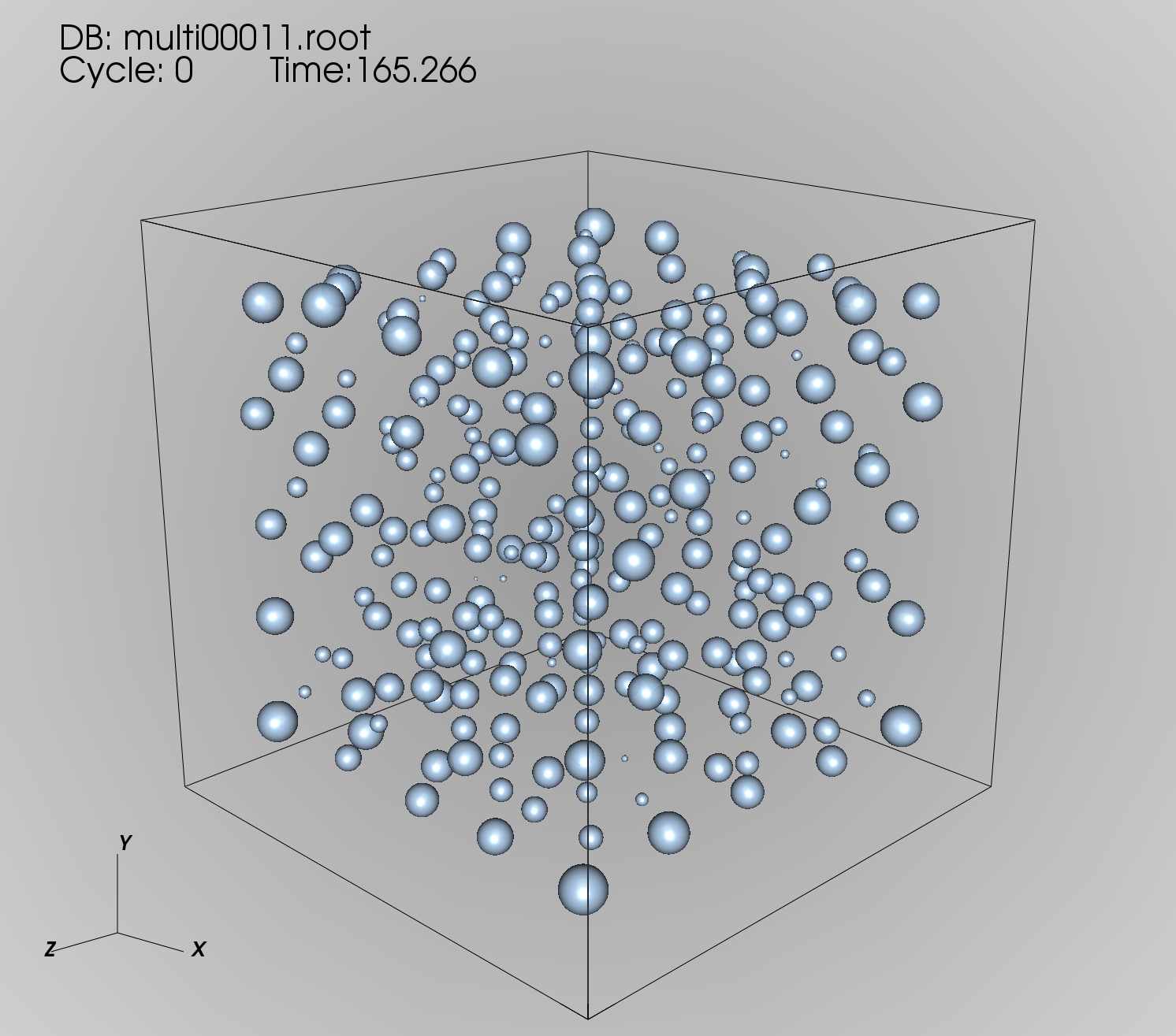} &
\includegraphics[width=.5\textwidth]{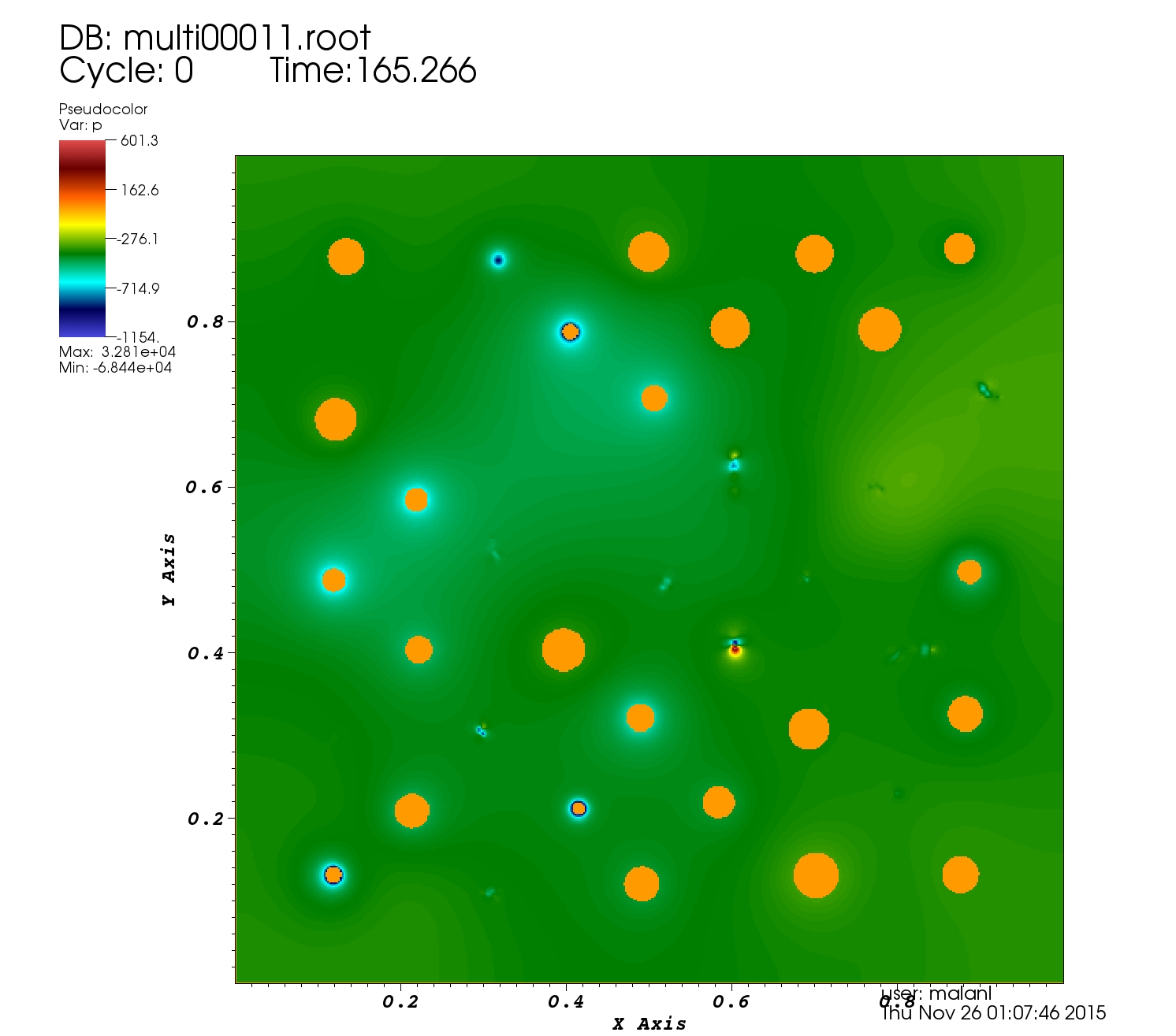} \\ 
\includegraphics[width=.5\textwidth]{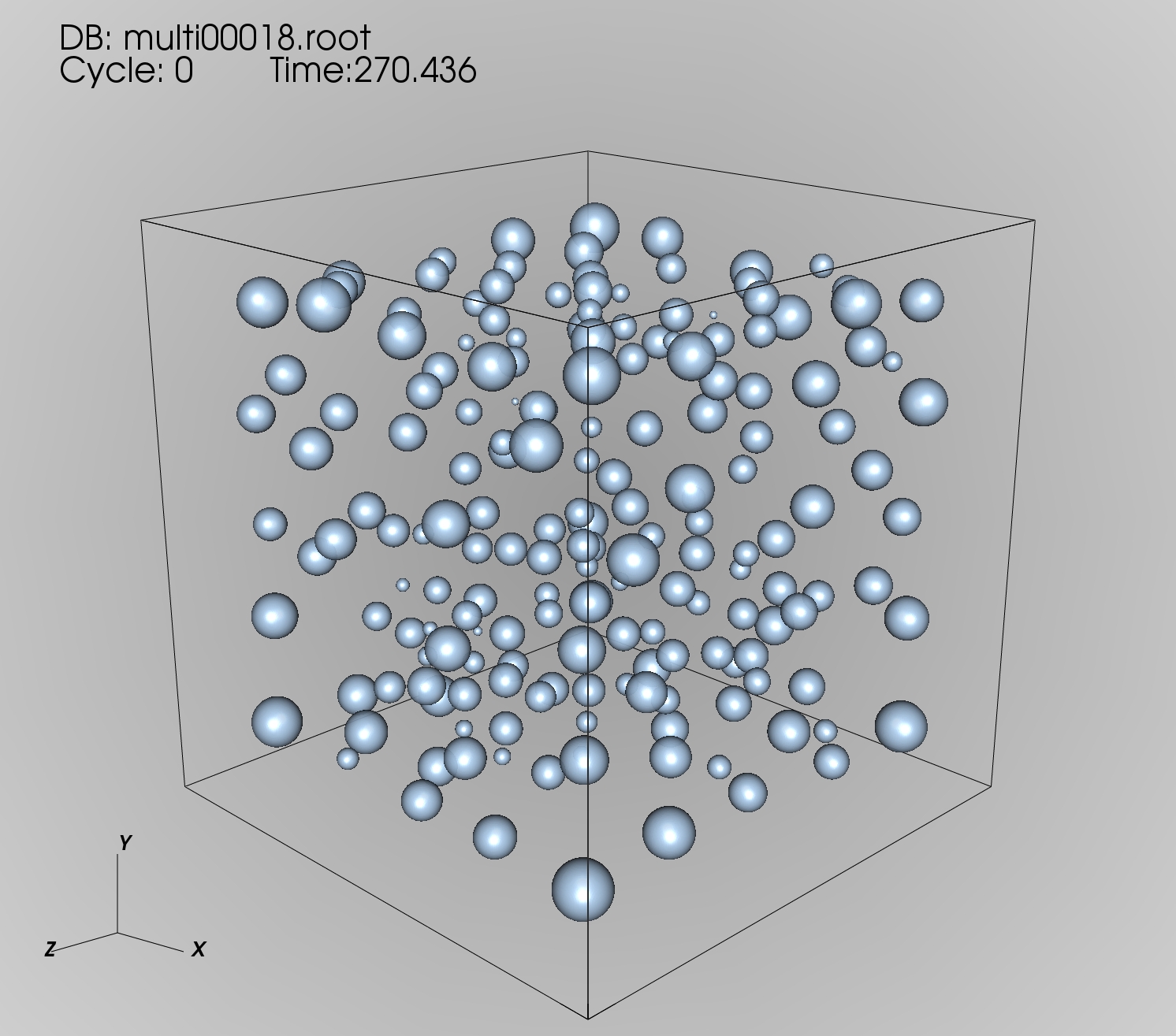} & 
\includegraphics[width=.5\textwidth]{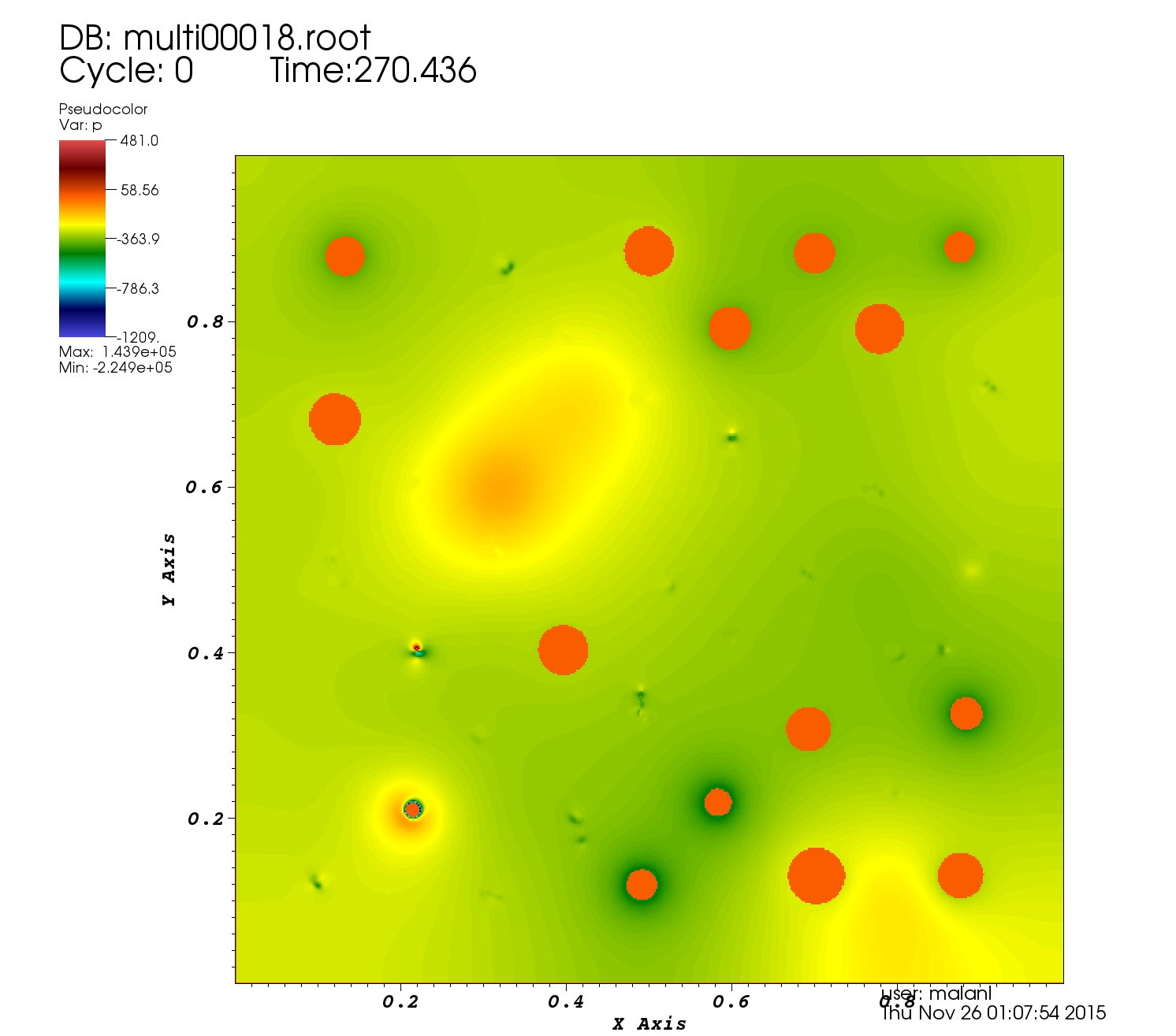}
\end{tabular}
\caption{VisIt screenshots of a simulation with 365 initial bubbles. The left shows a 3D view of bubbles at progressive time steps. The images on the right show the pressure distribution at the same instances for a section at $z = 0.5$. }
\label{fig:visit_snaps}
\end{figure*}

We observe that the higher the Weber number, the later bubble collapse occurs. The number of bubble collapses
at a given total void fraction decreases with increasing We. Fig. \ref{fig:visit_snaps} shows screen shots at progressive time steps. We observe the growth of some larger bubbles at the demise of smaller ones. The two-dimensional slices on the right show bubbles with a pressure heat map. With the vanishing vapour pressure model, a pressure gradient is formed in the liquid from large to small bubbles. This is the effect of surface tension.
%-----------------------------------------
%-----------------------------------------

\begin{figure*}
\begin{tabular}[width=0.95\textwidth]{ccc}
$We=5 \times 10^{-4}$ & $We = 1.3 \times 10^{-2}$ & $We=0.54$ \\
\includegraphics[width=.3\textwidth]{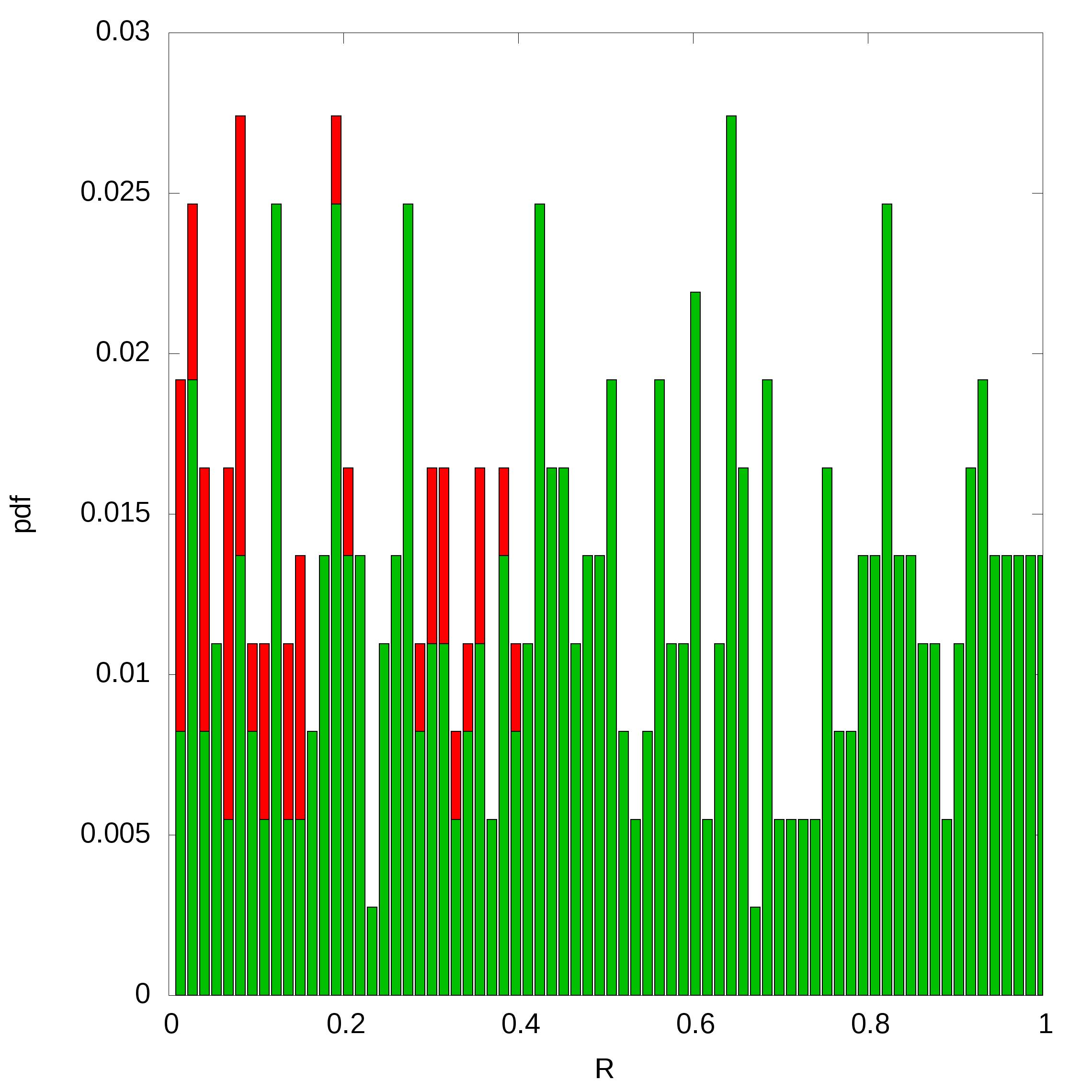} & 
\includegraphics[width=.3\textwidth]{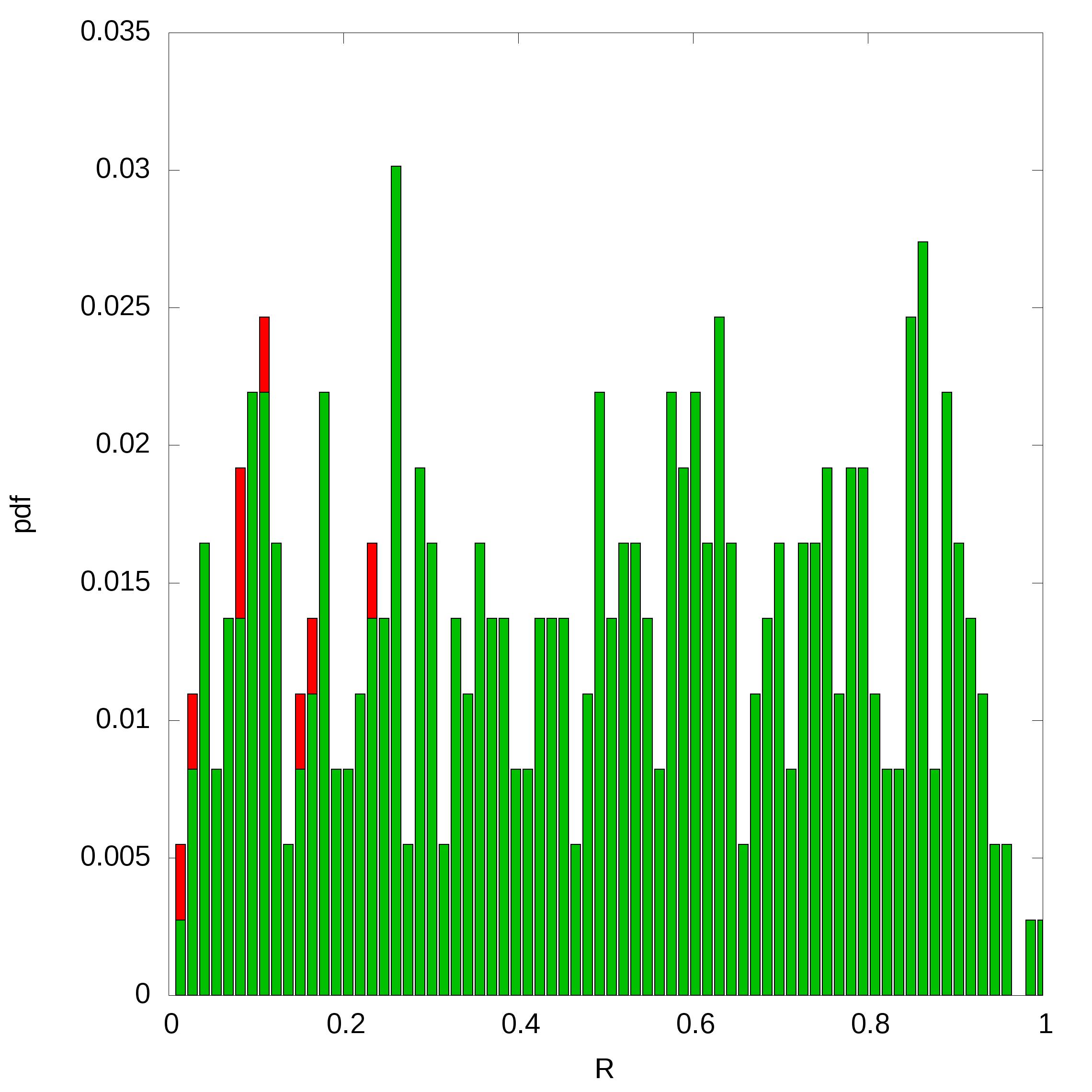} & 
\includegraphics[width=.3\textwidth]{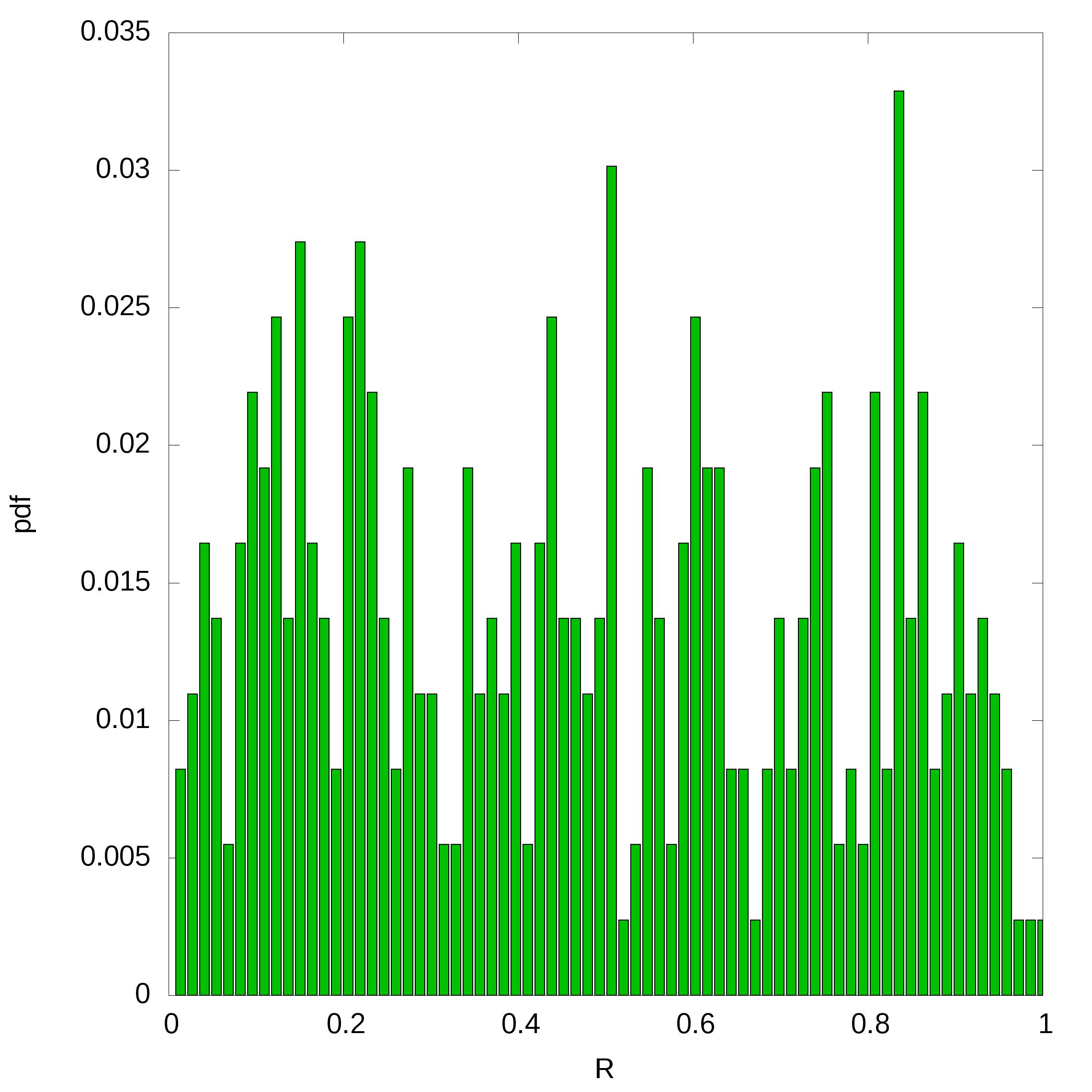}\\
\includegraphics[width=.3\textwidth]{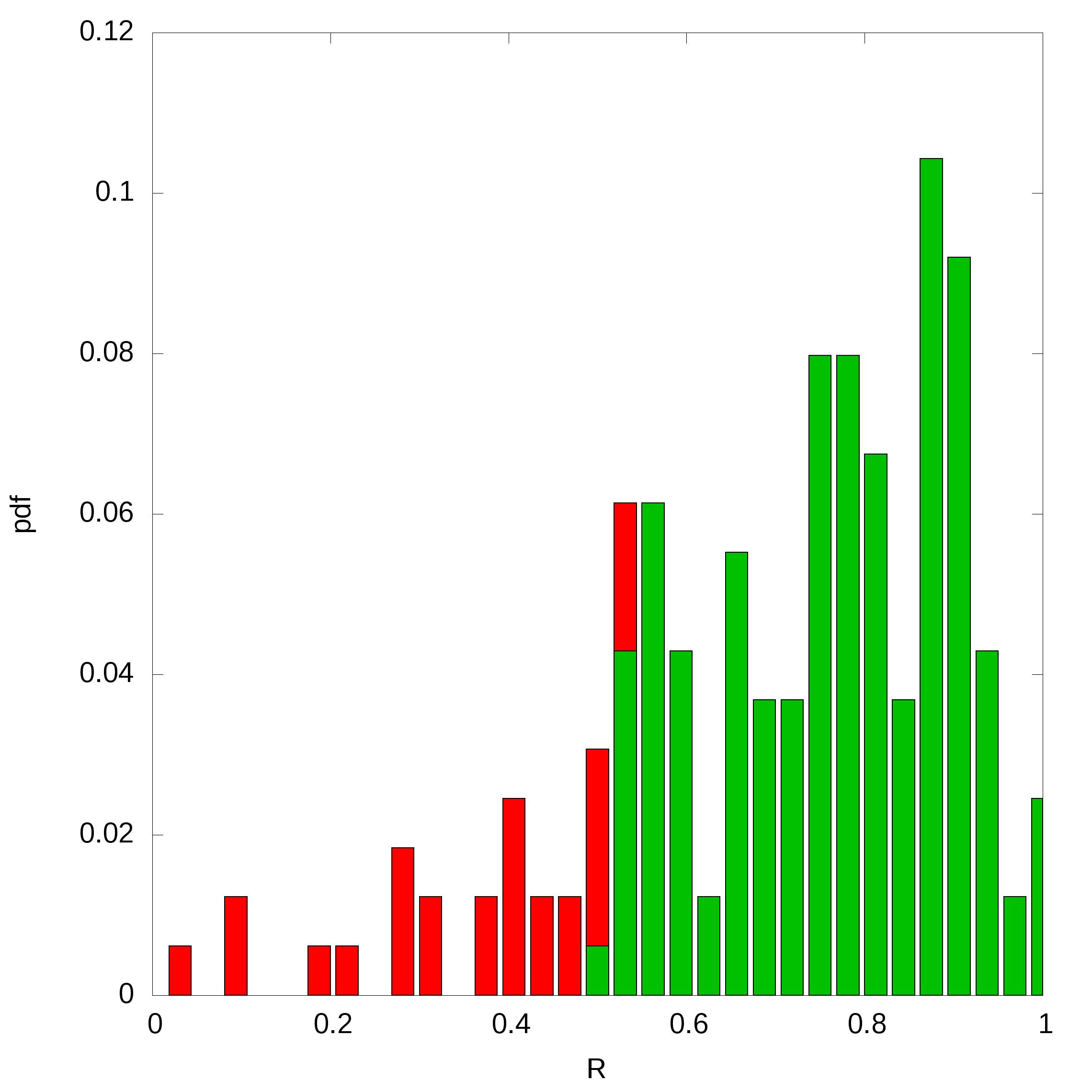} & 
\includegraphics[width=.3\textwidth]{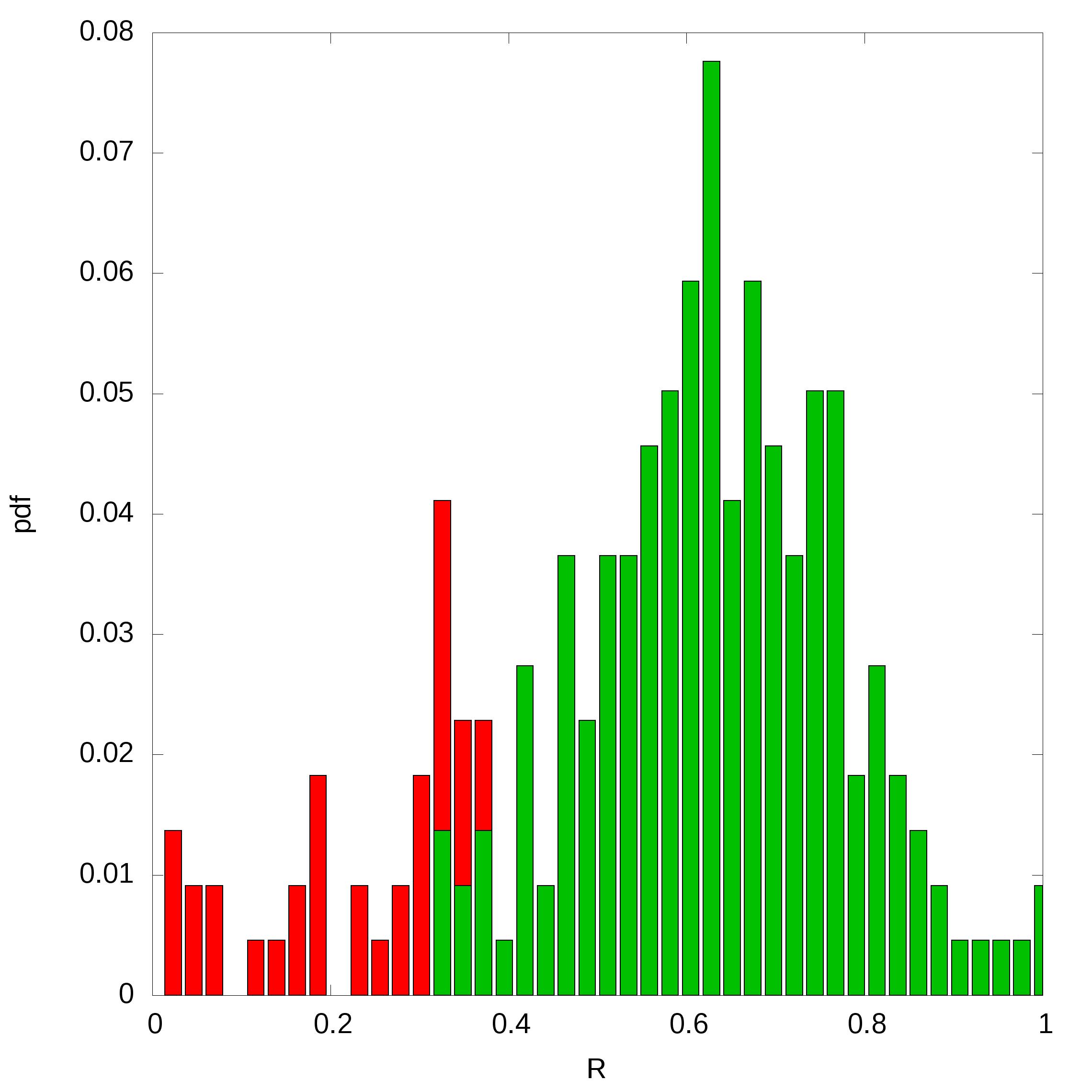} & 
\includegraphics[width=.3\textwidth]{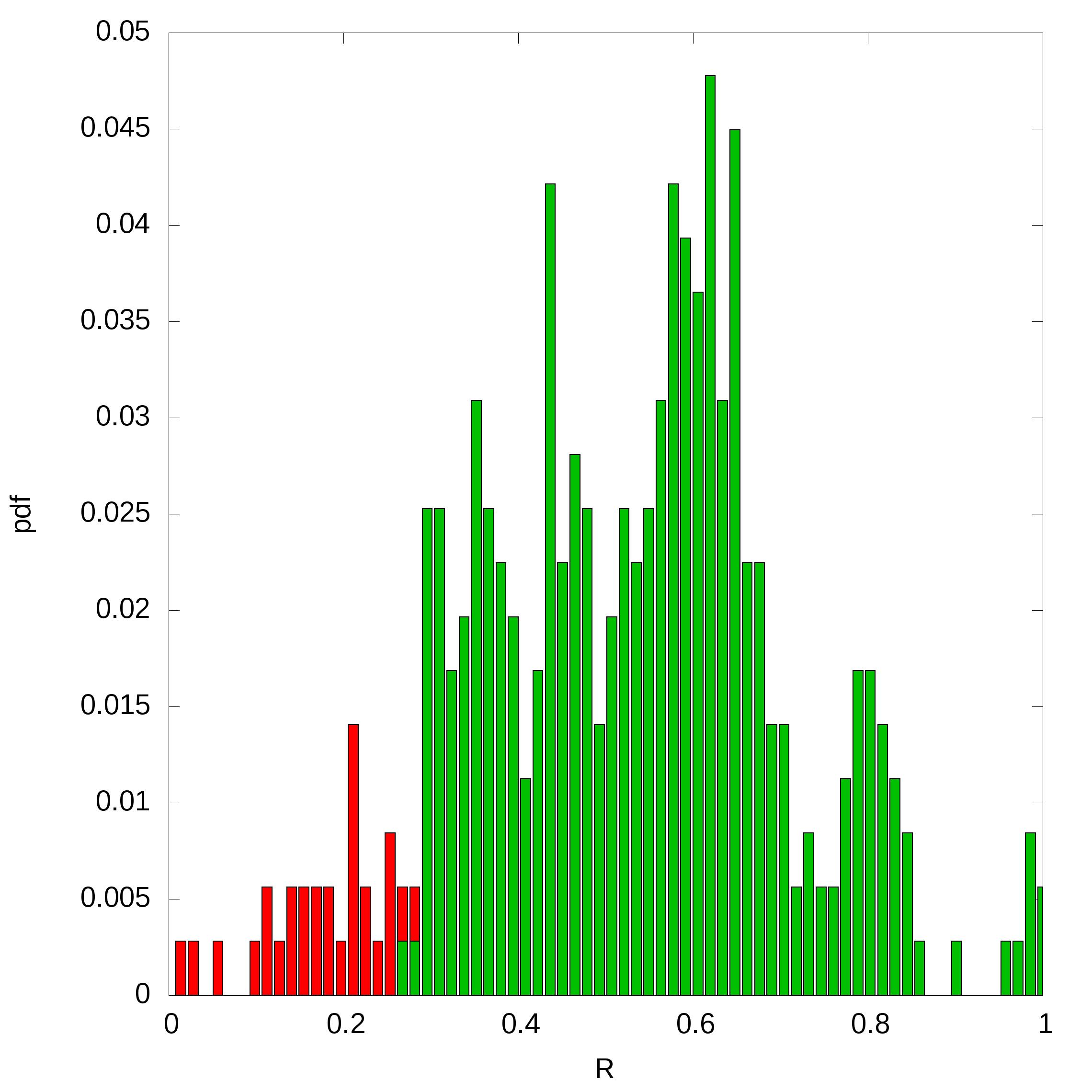}
\end{tabular}
\caption{Distribution of bubble radii, with growing bubbles shown in green. 
The first row shows the initial distribution and the second after 3000 time 
steps.}
\label{fig:pdfs_we}
\end{figure*}

The bubble radius distribution is presented in fig. \ref{fig:pdfs_we}. 
Initially all bubbles expand for the high We case.
%-----------------------------------------
%-----------------------------------------
\subsection{Bubble interaction}

\begin{figure}[h]
\centering
\includegraphics[width=0.7\textwidth]{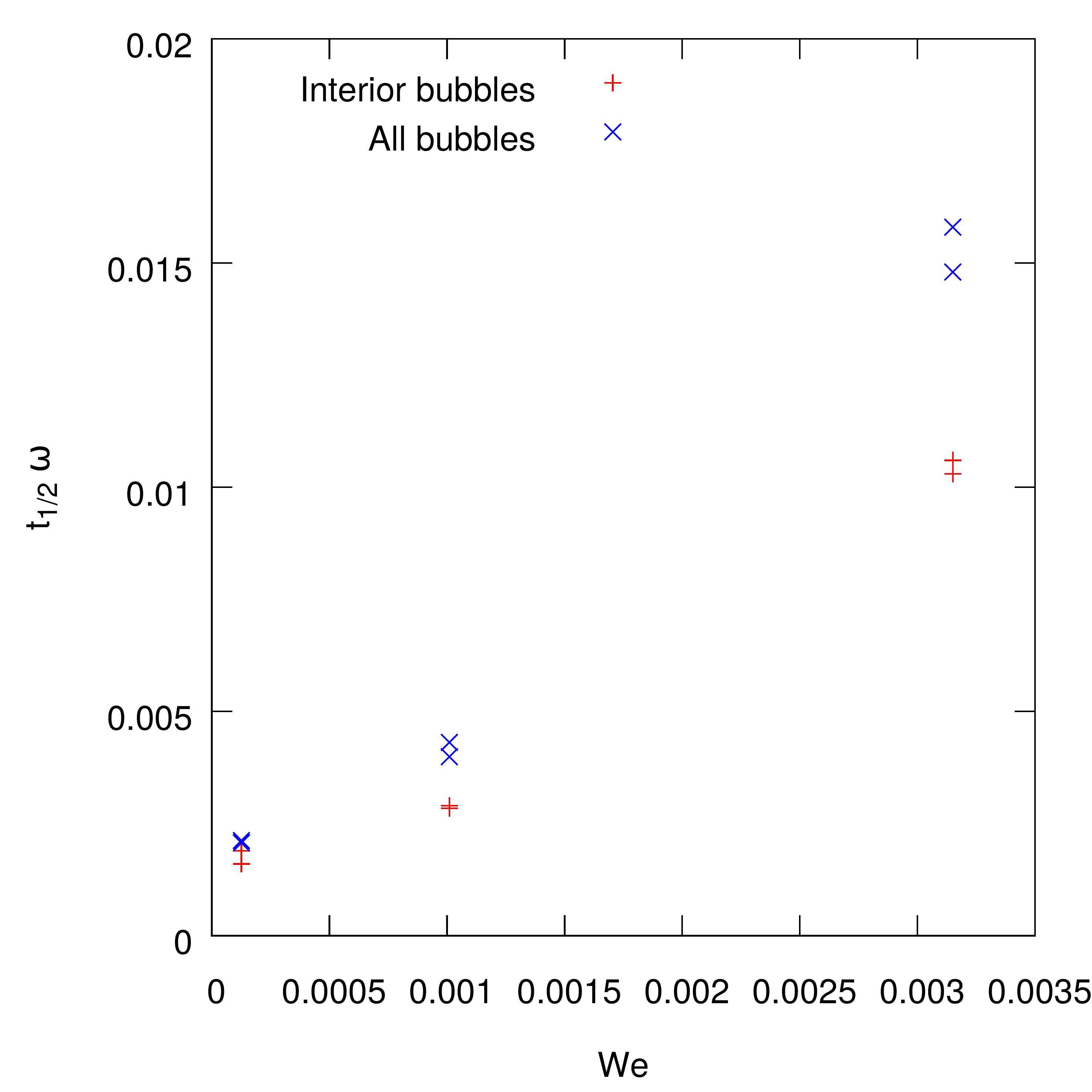}
\caption{Comparison of non-dimensional collapse time for half the 
initial bubbles, $t_{\rfrac{1}{2}} \omega$ for three We cases.}
\label{fig:tomega}
\end{figure}

In this section we formulate a proportionality between the time scale 
of pore competition and the Weber number. Since the liquid is incompressible, 
the fluid outflow with a constant velocity is balanced by an overall
volume expansion of the bubbles in the computational domain
%-----------------------------------------
\begin{equation}\label{eq:volume_growth}
\omega L^{3} = \sum_{i=1}^{N} 4 \pi R_{i}^{2} \dot{R_i}
\end{equation}
%-----------------------------------------
If we now assume that the sum of the volume expansion of each bubble
can be written in terms of the volume expansion of an average bubble
of radius $\overline{R}$ and average rate of change of its radius
$\dot{\overline{R}}$
%-----------------------------------------
\begin{equation}
\sum_{i=1}^{N} 4 \pi R_{i}^{2} \dot{R_i} = N 4 \pi \overline{R}^{2} 
\dot{\overline{R}}
\end{equation}
%-----------------------------------------
we can integrate \eqref{eq:volume_growth} to obtain the average radius 
evolution
%-----------------------------------------
\begin{equation}\label{eq:R_avg}
\overline{R}(t)^{3} = \dfrac{3 \omega L^{3}t}{4 \pi N}
\end{equation}
%-----------------------------------------
assuming $R_{0} \ll 1$. This is true since bubbles are initialised to be as small as possible. Let $t_{\rfrac{1}{2}}$ be the 
time at which half of the bubbles have collapsed, then we can now write 
%-----------------------------------------
\begin{equation}
t_{\rfrac{1}{2}} = \left( \dfrac{\rho \overline{R}
(t_{\rfrac{1}{2}})^{3}}{\sigma}\right)^{1/2}
\end{equation}
%-----------------------------------------
If we substitute \eqref{eq:R_avg} and rearrange we get 
%-----------------------------------------
\begin{equation}
t_{\rfrac{1}{2}} \omega = \dfrac{3\omega^{2}\rho \,\ell_{D}^{3}}{4 \pi \sigma} 
= \dfrac{3}{4 \pi} We_{\ell_{D}}
\end{equation}
%-----------------------------------------
This was tested for the cases presented in the previous section and the 
results are given in Fig. \ref{fig:tomega}.
%-----------------------------------------
%-----------------------------------------
The time $t_{\rfrac{1}{2}}$ was measured by considering all the bubbles 
inside the domain or by excluding the outermost ones. 
It is interesting to note that the measured times differ, especially 
for the higher We case. This indicates a buffering effect exerted by the 
outermost bubbles leading to different evolution rates for the bubbles 
towards the interior. The relationship is at least qualitatively linear, 
but should be confirmed with tests at a wide range of Weber numbers.

%-----------------------------------------
\section*{Conclusion}

A numerical tool was presented to deal with the specific problem of bubble 
interaction in an expanding, incompressible liquid during the micro-spalling 
of metals. The code was validated by simulating a single oscillating bubble 
and comparing it to the Rayleigh-Plesset solution. 

Simulations with hundreds of bubbles were studied for test cases defined by a 
Weber number. A bubble competition was observed and the time scale of bubble 
interaction was found to be dependent on the Weber number.

%-----------------------------------------

\section*{Acknowledgement}
This project was funded by the CEA and simulations were performed on their 
\textit{Airain} supercomputer at the \textit{TGCC} facility. We thank them 
for their kind co-operation.

Visualization and images, like those in Fig. \ref{fig:visit_snaps} were created using \textit{VisIt} \cite{visit}. \textit{VisIt} is supported by the Department of Energy with funding from the Advanced Simulation and Computing Program and the Scientific Discovery through Advanced Computing Program. 

Thank you also to Wojciech Aniszewski for valuable feedback during the writing of this paper.

\bibliographystyle{elsarticle-num}
\bibliography{BubCloudBib}

\begin{thebibliography}{10}
\expandafter\ifx\csname url\endcsname\relax
  \def\url#1{\texttt{#1}}\fi
\expandafter\ifx\csname urlprefix\endcsname\relax\def\urlprefix{URL }\fi
\expandafter\ifx\csname href\endcsname\relax
  \def\href#1#2{#2} \def\path#1{#1}\fi

\bibitem{2008SignorThesis}
L.~Signor, Contribution {\`{a}} la caract{\'{e}}risation et {\`{a}} la
  mod{\'{e}}lisation du micro-{\'{e}}caillage de l{'}{\'{e}}tain fondu sous
  choc, Ph.D. thesis, ISAE-ENSMA Ecole Nationale Sup{\'{e}}rieure de
  M{\'{e}}canique et d{'}A{\'{e}}rotechique-Poitiers (2008).

\bibitem{2008Signor}
L.~Signor, A.~Dragon, G.~Roy, T.~De~Ress{\'{e}}guier, F.~Llorca, Dynamic
  fragmentation of melted metals upon intense shock wave loading. some
  modelling issues applied to a tin target, Archives of Mechanics 60~(4) (2008)
  323--343.

\bibitem{2009DeResseguier}
T.~De~Ress{\'{e}}guier, E.~Lescoute, G.~Morard, F.~Guyot, Dynamic fragmentation
  as a possible diagnostic for high pressure melting in laser shock-loaded
  iron, Shock Compression of Condensed Matter 2009: Proceedings of the American
  Physical Society Topical Group on Shock Compression of Condensed Matter. AIP
  Conference Proceedings 1195~(1) (2009) 1007--1010.
\newblock \href {http://dx.doi.org/10.1063/1.3294969}
  {\path{doi:10.1063/1.3294969}}.

\bibitem{2010DeResseguier}
T.~De~Ress{\'{e}}guier, D.~Loison, E.~Lescoute, L.~Signor, A.~Dragon, Dynamic
  fragmentation of laser shock-melted metals: Some experimental advances,
  Journal Of Theoretical And Applied Mechanics 48 (2010) 957--972.
\newblock \href
  {http://dx.doi.org/http://www.ptmts.org.pl/jtam/index.php/jtam/article/view/v48n4p957}
  {\path{doi:http://www.ptmts.org.pl/jtam/index.php/jtam/article/view/v48n4p957}}.

\bibitem{2010aSignor}
L.~Signor, E.~Lescoute, D.~Loison, T.~De~Ress{\'{e}}guier, A.~Dragon, G.~Roy,
  Experimental study of dynamic fragmentation of shockloaded metals below and
  above melting, ICEM 14 - 14th International Conference on Experimental
  Mechanics, Poitiers, France, Edited by Fabrice Br{\'{e}}mand; EPJ Web of
  Conferences 6 (2010) 39012.
\newblock \href {http://dx.doi.org/10.1051/epjconf/20100639012}
  {\path{doi:10.1051/epjconf/20100639012}}.

\bibitem{2014Fuster}
D.~Fuster, K.~Pham, S.~Zaleski, Stability of bubbly liquids and its connection
  to the process of cavitation inception, Physics of Fluids 26~(4) (2014)
  042002.
\newblock \href {http://dx.doi.org/10.1063/1.4870820}
  {\path{doi:10.1063/1.4870820}}.

\bibitem{2010bSignor}
L.~Signor, T.~De~Ress{\'{e}}guier, A.~Dragon, G.~Roy, A.~Fanget, M.~Faessel,
  \href{//www.sciencedirect.com/science/article/pii/S0734743X10000424}{Investigation
  of fragments size resulting from dynamic fragmentation in melted state of
  laser shock-loaded tin}, International Journal of Impact Engineering 37~(8)
  (2010) 887--900.
\newblock \href {http://dx.doi.org/10.1016/j.ijimpeng.2010.03.001}
  {\path{doi:10.1016/j.ijimpeng.2010.03.001}}.
\newline\urlprefix\url{//www.sciencedirect.com/science/article/pii/S0734743X10000424}

\bibitem{2001Stebnovskii}
S.~V. Stebnovskii, \href{http://dx.doi.org/10.1023/A:1019254906248}{Generalized
  rheological model of cavitating condensed media}, Journal of Applied
  Mechanics and Technical Physics 42~(3) (2001) 482--492.
\newblock \href {http://dx.doi.org/10.1023/A:1019254906248}
  {\path{doi:10.1023/A:1019254906248}}.
\newline\urlprefix\url{http://dx.doi.org/10.1023/A:1019254906248}

\bibitem{1999Scardovelli}
R.~Scardovelli, S.~Zaleski, Direct numerical simulation of free-surface and
  interfacial flow, Annual Review of Fluid Mechanics 31 (1999) 567--603.
\newblock \href {http://dx.doi.org/10.1146/annurev.fluid.31.1.567}
  {\path{doi:10.1146/annurev.fluid.31.1.567}}.

\bibitem{2006Tryggvason}
G.~Tryggvason, A.~Esmaeeli, J.~Lu, S.~Biswas,
  \href{http://www.sciencedirect.com/science/article/pii/S0169598306000554}{Direct
  numerical simulations of gas/liquid multiphase flows}, Fluid Dynamics
  Research 38~(9) (2006) 660--681.
\newblock \href {http://dx.doi.org/10.1016/j.fluiddyn.2005.08.006}
  {\path{doi:10.1016/j.fluiddyn.2005.08.006}}.
\newline\urlprefix\url{http://www.sciencedirect.com/science/article/pii/S0169598306000554}

\bibitem{2011Book}
G.~Tryggvason, R.~Scardovelli, S.~Zaleski,
  \href{https://books.google.co.za/books?id=nY5bjSYq-AEC}{Direct Numerical
  Simulations of Gas{--}Liquid Multiphase Flows}, Cambridge University Press,
  2011.
\newline\urlprefix\url{https://books.google.co.za/books?id=nY5bjSYq-AEC}

\bibitem{2005Tryggvason}
G.~Tryggvason, A.~Esmaeeli, N.~Al-Rawahi,
  \href{http://www.sciencedirect.com/science/article/pii/S0045794904004158}{Direct
  numerical simulations of flows with phase change}, Computers \& Structures
  83~(6{--}7) (2005) 445--453.
\newblock \href {http://dx.doi.org/10.1016/j.compstruc.2004.05.021}
  {\path{doi:10.1016/j.compstruc.2004.05.021}}.
\newline\urlprefix\url{http://www.sciencedirect.com/science/article/pii/S0045794904004158}

\bibitem{1965Harlow}
F.~H. Harlow, J.~E. Welch, Numerical calculation of time-dependent viscous
  incompressible flow of fluid with free surface, Physics of Fluids 8~(12)
  (1965) 2182--2189.
\newblock \href {http://dx.doi.org/10.1063/1.1761178}
  {\path{doi:10.1063/1.1761178}}.

\bibitem{2002Popinet}
S.~Popinet, S.~Zaleski, Bubble collapse near a solid boundary: a numerical
  study of the influence of viscosity, Journal of Fluid Mechanics 464~(01)
  (2002) 137--163.
\newblock \href {http://dx.doi.org/10.1017/S002211200200856X}
  {\path{doi:10.1017/S002211200200856X}}.

\bibitem{2016Tan}
H.~Tan,
  \href{http://www.sciencedirect.com/science/article/pii/S0301932215300434}{An
  adaptive mesh refinement based flow simulation for free-surfaces in thermal
  inkjet technology}, International Journal Of Multiphase Flow 82 (2016) 1--16.
\newblock \href {http://dx.doi.org/016/j.ijmultiphaseflow.2016.01.001}
  {\path{doi:016/j.ijmultiphaseflow.2016.01.001}}.
\newline\urlprefix\url{http://www.sciencedirect.com/science/article/pii/S0301932215300434}

\bibitem{1993Szymczak}
W.~G. Szymczak, J.~C.~W. Rogers, J.~M. Solomon, A.~E. Bergert,
  \href{http://www.sciencedirect.com/science/article/pii/S0021999183711113}{A
  numerical algorithm for hydrodynamic free boundary problems}, Journal of
  Computational Physics 106~(2) (1993) 319--336.
\newblock \href {http://dx.doi.org/10.1016/S0021-9991(83)71111-3}
  {\path{doi:10.1016/S0021-9991(83)71111-3}}.
\newline\urlprefix\url{http://www.sciencedirect.com/science/article/pii/S0021999183711113}

\bibitem{1981Hirt}
C.~W. Hirt, B.~D. Nichols,
  \href{http://www.sciencedirect.com/science/article/pii/0021999181901455}{Volume
  of fluid (vof) method for the dynamics of free boundaries}, Journal of
  Computational Physics 39~(1) (1981) 201--225.
\newblock \href {http://dx.doi.org/10.1016/0021-9991(81)90145-5}
  {\path{doi:10.1016/0021-9991(81)90145-5}}.
\newline\urlprefix\url{http://www.sciencedirect.com/science/article/pii/0021999181901455}

\bibitem{1969Chorin}
A.~J. Chorin, On the convergence of discrete approximations to the
  navier-stokes equations, Mathematics of computation 23~(106) (1969) 341--353.

\bibitem{1974DeBar}
R.~DeBar, Fundamentals of the kraken code, Lawrence Livermore Laboratory,
  UCIR-760.

\bibitem{2007Aulisa}
E.~Aulisa, S.~Manservisi, R.~Scardovelli, S.~Zaleski, Interface reconstruction
  with least-squares fit and split advection in three-dimensional cartesian
  geometry, Journal of Computational Physics 225~(2) (2007) 2301--2319.
\newblock \href {http://dx.doi.org/10.1016/j.jcp.2007.03.015}
  {\path{doi:10.1016/j.jcp.2007.03.015}}.

\bibitem{2000Scardovelli}
R.~Scardovelli, S.~Zaleski, Analytical relations connecting linear interfaces
  and volume fractions in rectangular grids, Journal of Computational Physics
  164~(1) (2000) 228--237.
\newblock \href {http://dx.doi.org/10.1006/jcph.2000.6567}
  {\path{doi:10.1006/jcph.2000.6567}}.

\bibitem{1995Li}
J.~Li, Calcul d{'}interface affine par morceaux, Comptes rendus de
  l{'}Acad{\'{e}}mie des sciences. S{\'{e}}rie II, M{\'{e}}canique, physique,
  chimie, astronomie 320~(8) (1995) 391--396.

\bibitem{2010Weymouth}
G.~D. Weymouth, D.~K.-P. Yue, Conservative volume-of-fluid method for
  free-surface simulations on cartesian-grids, Journal of Computational Physics
  229~(8) (2010) 2853--2865.
\newblock \href {http://dx.doi.org/10.1016/j.jcp.2009.12.018}
  {\path{doi:10.1016/j.jcp.2009.12.018}}.

\bibitem{2004Esmaeeli}
A.~Esmaeeli, G.~Tryggvason,
  \href{http://www.sciencedirect.com/science/article/pii/S0017931004002947}{Computations
  of film boiling. part i: numerical method}, International Journal of Heat and
  Mass Transfer 47~(25) (2004) 5451--5461.
\newblock \href {http://dx.doi.org/10.1016/j.ijheatmasstransfer.2004.07.027}
  {\path{doi:10.1016/j.ijheatmasstransfer.2004.07.027}}.
\newline\urlprefix\url{http://www.sciencedirect.com/science/article/pii/S0017931004002947}

\bibitem{2003Sussman}
M.~Sussman, A second order coupled level set and volume-of-fluid method for
  computing growth and collapse of vapor bubbles, Journal of Computational
  Physics 187~(1) (2003) 110--136.
\newblock \href {http://dx.doi.org/10.1016/S0021-9991(03)00087-1}
  {\path{doi:10.1016/S0021-9991(03)00087-1}}.

\bibitem{2009Popinet}
S.~Popinet, An accurate adaptive solver for surface-tension-driven interfacial
  flows, Journal Of Computational Physics 228 (2009) 5838--5866.
\newblock \href {http://dx.doi.org/016/j.jcp.2009.04.042}
  {\path{doi:016/j.jcp.2009.04.042}}.

\bibitem{1999Fedkiw}
R.~P. Fedkiw, T.~Aslam, B.~Merriman, S.~Osher, A non-oscillatory eulerian
  approach to interfaces in multimaterial flows (the ghost fluid method),
  Journal of Computational Physics 152~(2) (1999) 457--492.
\newblock \href {http://dx.doi.org/10.1006/jcph.1999.6236}
  {\path{doi:10.1006/jcph.1999.6236}}.

\bibitem{2000Kang}
M.~Kang, R.~P. Fedkiw, X.-D. Liu,
  \href{http://dx.doi.org/10.1023/A:1011178417620}{A boundary condition
  capturing method for multiphase incompressible flow}, Journal of Scientific
  Computing 15~(3) (2000) 323--360.
\newblock \href {http://dx.doi.org/10.1023/A:1011178417620}
  {\path{doi:10.1023/A:1011178417620}}.
\newline\urlprefix\url{http://dx.doi.org/10.1023/A:1011178417620}

\bibitem{2010Herrmann}
M.~Herrmann, A parallel eulerian interface tracking/lagrangian point particle
  multi-scale coupling procedure, Journal of Computational Physics 229~(3)
  (2010) 745--759.
\newblock \href {http://dx.doi.org/10.1016/j.jcp.2009.10.009}
  {\path{doi:10.1016/j.jcp.2009.10.009}}.

\bibitem{1970Chan}
R.~K.-C. Chan, R.~L. Street, A computer study of finite-amplitude water waves,
  Journal of Computational Physics 6~(1) (1970) 68--94.
\newblock \href {http://dx.doi.org/10.1016/0021-9991(70)90005-7}
  {\path{doi:10.1016/0021-9991(70)90005-7}}.

\bibitem{1979Leonard}
B.~P. Leonard, A stable and accurate convective modelling procedure based on
  quadratic upstream interpolation, Computer Methods in Applied Mechanics and
  Engineering 19 (1979) 59--98.
\newblock \href {http://dx.doi.org/10.1016/0045-7825(79)90034-3}
  {\path{doi:10.1016/0045-7825(79)90034-3}}.

\bibitem{1986Harten}
A.~HARTEN, S.~OSHER, B.~ENGQUIST, S.~CHAKRAVARTHY, Some results on uniformly
  high-order accurate essentially nonoscillatory schemes, Applied Numerical
  Mathematics 2 (1986) 347--377.
\newblock \href {http://dx.doi.org/10.1016/0168-9274(86)90039-5}
  {\path{doi:10.1016/0168-9274(86)90039-5}}.

\bibitem{2009Shu}
C.-W. Shu, High order weighted essentially nonoscillatory schemes for
  convection dominated problems, Siam Review 51 (2009) 82--126.
\newblock \href {http://dx.doi.org/137/070679065} {\path{doi:137/070679065}}.

\bibitem{1977Plesset}
M.~S. Plesset, A.~Prosperetti, Bubble dynamics and cavitation, Annual review of
  fluid mechanics 9~(1) (1977) 145--185.

\bibitem{2016Bna}
S.~Bn{\`{a}}, S.~Manservisi, R.~Scardovelli, P.~Yecko, S.~Zaleski, Vofi - a
  library to initialize the volume fraction scalar field, Computer Physics
  Communications 200 (2016) 291--299.
\newblock \href {http://dx.doi.org/10.1016/j.cpc.2015.10.026}
  {\path{doi:10.1016/j.cpc.2015.10.026}}.

\bibitem{visit}
H.~Childs, E.~Brugger, B.~Whitlock, J.~Meredith, S.~Ahern, D.~Pugmire,
  K.~Biagas, M.~Miller, C.~Harrison, G.~H. Weber, H.~Krishnan, T.~Fogal,
  A.~Sanderson, C.~Garth, E.~W. Bethel, D.~Camp, O.~R{\"{u}}bel, M.~Durant,
  J.~M. Favre, P.~Navr{\'{a}}til, Visit: An end-user tool for visualizing and
  analyzing very large data, in: High Performance Visualization{--}Enabling
  Extreme-Scale Scientific Insight, 2012, pp. 357--372.

\end{thebibliography}
\end{document}